\documentclass[preprint,12pt]{elsarticle}

\usepackage{graphicx,psfrag,color} 
  \graphicspath{ {./images/} }

  \usepackage{adjustbox}

\usepackage{algorithm}
\usepackage[noend]{algpseudocode}

\usepackage{caption}

 \usepackage{epstopdf}
\usepackage{amssymb}
\usepackage{amsmath}

\usepackage{subcaption}

\begin{document}

\begin{frontmatter}









\author[1]{Meixuan Lin\corref{cor1}} 
\ead{meixuan.lin@postgrad.manchester.ac.uk}

\author[1]{Georgios Fourtakas} 
\author[1]{Benedict D. Rogers} 

\address[1]{Faculty of Science and Engineering, University of Manchester, Manchester, M13 9PL, UK}

\cortext[cor1]{Corresponding author}


\title{A high-order Fourier Continuation (FC)-based spectral incompressible Smoothed Particle
Hydrodynamics (ISPH) scheme for general boundary conditions in wall-bounded domains }






\begin{abstract}

In this paper, a high-order Fourier Continuation (FC) algorithm is introduced into the spectral smoothed particle hydrodynamics (SPH) scheme to simulate the wall-bounded incompressible flows. This work aims to extend the spectral ISPH scheme towards the high-order simulation of flows with non-periodic wall boundary conditions. Herein, a polynomial-based Fourier continuation technique is applied to the velocity and pressure to make the domain both periodic and $C^p$ smooth. The spatial SPH discretisation is performed subsequently in the frequency space on the FC-extended domain by building upon the convolution theorem using fast Fourier transform (FFT). The incorporation of Neumann boundary conditions is straightforward, and more generally, the FC method enforces periodicity across the domain regardless of the boundary condition type. The convergence order, additional computational cost, and implementation technique of the FC method are also discussed.
Combined with a projection-based time integration scheme and a spectral PPE solver, the FC-based spectral ISPH framework is validated against several classical CFD benchmarks. The principal finding of this work is that the incorporation of FC techniques enables the spectral ISPH scheme to simulate wall-bounded flows with high-order convergence, and accurately capturing complex vortex dynamics. This work therefore represents a step towards a fully high-order spectral Lagrangian SPH solver with complex geometries

\end{abstract}





\end{frontmatter}

\section{Introduction}



Computational fluid dynamics (CFD) methods can be broadly categorised into mesh-based and meshfree approaches. Smoothed Particle Hydrodynamics (SPH) is one of the most widely adopted meshfree methods, due to its Lagrangian nature in which the fluid is represented by a set of particles where the governing equations are discretised through kernel-based interpolation. Originally proposed for astrophysical simulations~\cite{PRICE2012759SPHMHD, phantom}, SPH has been extended to a wide range of engineering applications, such as multi-phase flows~\cite{MONAGHAN1995225Multiphase, HU2006JCPMultiphrase, CHEN2015169Multiphase}. The absence of a mesh allows particles to move freely along streamlines. This makes SPH particularly well-suited to problems involving large deformations and complex free-surface dynamics. 

Enabling high-order and computationally efficient SPH schemes is identified as one of the five grand challenges among the SPH community \cite{grandChallenge}. Since high-order schemes can achieve superior accuracy per degree of freedom compared to their low-order counterparts, thus making them more computationally efficient for resolving complex flow physics~\cite{Nektar++}. Based on the reviews of~\cite{2022RoyalReview} and~\cite{Meng2025HighOrderSPH}, state-of-the-art high-order SPH methods can be broadly classified into three categories: correcting SPH derivative operators to restore consistency conditions~\cite{SIBILLAM2,NASARHOCSPH}, constructing high-order kernel functions with improved spectral resolution~\cite{LIND2016JCP, NASAR2021HOWallBC,WANG2024113385}, and adopting spatial reconstruction within the Riemann-SPH framework to achieve higher-order flux accuracy~\cite{GAO2023112270,MENG20242024117065,Oger2023WENOSPH}. 

All of the aforementioned developments operate within the traditional physical-space SPH discretisation. In contrast, our previous work~\cite{LINSpectralSPH} introduced a fundamentally different approach by embedding high-order spectral methods~\cite{Trefethen} directly into the SPH framework. The SPH operators are reformulated in the frequency domain via the convolution theorem. This spectral feature enables efficient operator evaluation. Through the use of the fast Fourier transform (FFT), the computational complexity is reduced to $\mathcal{O}(N \log N)$, where $N$ denotes the number of SPH particles. Up to $6$th order of convergence rate and computational gains were demonstrated in \cite{LINSpectralSPH}, while most of the test cases are restricted into periodic domains due to the periodicity constraints of FFT. High-order wall boundary conditions represent a key limitation of the existing spectral ISPH scheme. To address this, a polynomial-based Fourier continuation (FC) method is introduced in this work to enable the framework to handle such conditions in a high-order manner. Fourier continuation is an algorithm originally proposed by Bruno et al.~\cite{FCADI, Albin2011FCNavierStokes,LYON20103358} to obtain accurate Fourier extensions of functions defined on non-periodic domains, through the use of FC(Gram) basis functions. The FC scheme efficiently removes the Gibbs phenomenon, recovers spectral accuracy and maintains the dispersionless property for general mesh-based spectral methods. This makes FC a natural fit for the spectral ISPH framework: by extending the solution periodically across domain boundaries, the existing spectral SPH discretisation can be applied directly on top of the continuation. This enables high-order treatment of arbitrary wall boundary conditions within a periodic spectral setting. 

In this work, a polynomial-based FC method is incorporated into the spectral ISPH framework. Through high-order polynomial fitting, boundary extrapolation and smooth blending, a $C^p$ periodic extension of length $d$ is constructed at each domain boundary, where $p$ and $d$ are user-defined parameters governing the smoothness and length of the continuation. The FC-based spectral ISPH scheme is combined with a spectral pressure Poisson equation (PPE) solver to enforce incompressibility. Several validation cases are implemented to show the potential of the scheme in satisfying inflow--outflow, general Dirichlet and Neumann boundary conditions, and in accurately resolving the vortex--wall interaction of different Reynolds numbers.

The remainder of this paper is structured as follows. The governing incompressible Navier--Stokes equations and time integration schemes are introduced in Section~\ref{2}. Section~\ref{spectralSPHScheme} reviews the spectral SPH discretisation from our previous work and discusses the resolving power of various SPH kernels. The polynomial-based FC algorithm for imposing Dirichlet and Neumann boundary conditions is detailed in Section~\ref{sec:fc:1d}. Section~\ref{wholeframework} then describes the complete FC-based spectral ISPH framework, including the spectral PPE solver and the projection-based time-stepping procedure. Numerical validation against several classical CFD benchmarks is carried out in Section~\ref{Testcasesvalidation}, including a range of boundary condition types. Finally, Section~\ref{Conclusion} summarises the main findings and draws conclusions.

\section{General framework of the numerical method }\label{2}
\subsection{Governing equation} \label{2.1}
In this work, we consider the two-dimensional motion of a viscous incompressible fluid, described by the momentum and continuity equations:
\begin{equation}\label{IncompressibleNS}
\frac{\partial \textbf{u}}{\partial t}+(\textbf{u} \cdot\nabla) \textbf{u}=-\frac{1}{\rho}\nabla P+\nu\nabla^2 \textbf{u} + \textbf{f},
\end{equation}
\begin{equation}\label{divergenceFree}
\nabla \cdot \textbf{u}=0,
\end{equation}
where $\textbf{u}=[u,v]$ is the velocity vector, $P$ is the pressure, $\rho$ denotes the fluid density, $\nu$ the kinematic viscosity, and $\textbf{f}$ represents for any external forcing applied to the fluid. Dirichlet boundary condition is imposed to velocity and a homogeneous Neumann boundary condition is applied to pressure on the domain boundary $\partial\Omega$:
\begin{equation}\label{BC_velocity}
\textbf{u} = c \quad \text{on} \quad \partial\Omega,
\end{equation}
\begin{equation}\label{BC_pressure}
\frac{\partial P}{\partial \textbf{n}} = 0 \quad \text{on} \quad \partial\Omega,
\end{equation}
where $\textbf{n}$ is the outward unit normal to the boundary.

\subsection{Temporal discretisation} \label{2.2}
To advance Equations~(\ref{IncompressibleNS}) and~(\ref{divergenceFree}) in time, we adopt the first-order projection method (FOP)~\cite{1999JCP,1968chorin} to decouple the velocity and pressure computations. An intermediate velocity field $\textbf{u}^*$ is obtained via
\begin{equation}\label{uStarRK3}
\textbf{u}^{*}=\textbf{u}^k-\Delta t \left(g_k \textbf{F}^k+z_k\textbf{F}^{k-1}-a_k\nabla{\tilde{p}^k}+a_k\tilde{\mathbf{f}}^{k+1}\right),
\end{equation}
where
\begin{equation} \label{rhsMomtem}
\textbf{F}^k=\textbf{u}^k \cdot\nabla \textbf{u}^k+\nu\nabla^2 \textbf{u}^k,
\end{equation}
and $\tilde{p}^k=\frac{1}{a_k\Delta t}\int_{t_k}^{t_{k+1}}p\,\mathrm{d}t$ and $\tilde{\mathbf{f}}^k=\frac{1}{a_k\Delta t}\int_{t_k}^{t_{k+1}}f\,\mathrm{d}t$ denote the time-averaged pressure and body force within each sub-step $k$, respectively. A pressure $\phi$ is then obtained by solving the pressure Poisson equation (PPE):
\begin{equation} \label{PPE}
\nabla^2 \phi =\frac{\rho}{\Delta t} a_k\nabla\cdot \textbf{u}^*.
\end{equation}
The velocity at the next sub-step is subsequently corrected by the pressure gradient:
\begin{equation} \label{u^n+1}
\textbf{u}^{k+1}=\textbf{u}^*-a_k\frac{\Delta t}{\rho}\nabla \phi.
\end{equation}
Two temporal discretisation schemes are employed in this study. The primary scheme is a low-storage third-order Runge--Kutta method (RK3), with the coefficients
\[
a_k=\left[\frac{8}{15},\,\frac{2}{15},\,\frac{1}{3}\right], \quad
g_k=\left[\frac{8}{15},\,\frac{5}{12},\,\frac{3}{4}\right], \quad
z_k=\left[0,\,-\frac{17}{60},\,-\frac{5}{12}\right].
\]
Setting $g_k = 1$, $z_k = 0$, and $a_k = 1$ reduces the scheme to the conventional first-order projection method.

\section{Spectral SPH discretisation scheme}
\label{spectralSPHScheme}
\subsection{Spectral SPH operators}
Classically, the SPH derivative operator at particle $i$ in physical space is discretised as follows by summing the contribution of all the particle $j$ located in the influence domain of $i$:
\begin{equation}\label{SPHConV}
\frac{\partial f(\textbf{x}_i)}{\partial \textbf{x}}=\sum_{j}f(\textbf{x}_j)\nabla_iW_{ij}V_j,
\end{equation}
where $j$ denotes all particles within the support radius of particle $i$ and $V_j$ is the particle volume. In our previous work \cite{LINSpectralSPH}, a novel spectral ISPH scheme was proposed that exploits the convolution theorem to reformulate operators
 to spectral space:
\begin{equation}\label{final_dft_gradient}
\frac{\partial f(\textbf{x}_i)}{\partial \textbf{x}}=\frac{1}{N^2} \sum^{N-1}_{k=0} \left[(\sum^{N-1}_{n=0}f_n e^{-i2\pi\frac{k}{N}n}) \cdot (\sum^{N-1}_{n=0}(\nabla W_n) e^{-i2\pi\frac{k}{N}n})\right]e^{i2\pi\frac{k}{N}n},
\end{equation}
where $N$ is the total number of particles in the computational domain and $k$ is the wavenumber index in the frequency domain. The fast Fourier transform (FFT) is employed to reduce the complexity from $\mathcal{O}(N^2)$ to $\mathcal{O}(NlogN)$, and the spectral SPH discretisation is applied in advection, viscous and $\nabla{\cdot \textbf{u}^*}$ and pressure gradient terms. Detailed analysis of the accuracy, efficiency and a combination of the scheme with the immersed boundary methods (IBM) can be found in \cite{LINSpectralSPH}.

\subsection{Linear and circular convolution}
\begin{figure}[h!]
    \centering
   
    \begin{subfigure}{.48\textwidth}
        \centering
         
        \includegraphics[width=\linewidth]{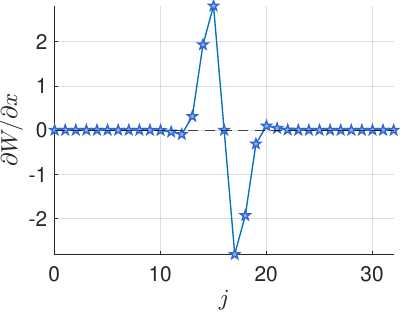}
        \caption{ kernel gradient $\frac{\partial W}{\partial x}$ }
    \end{subfigure}
    \begin{subfigure}{.48\textwidth}
        \centering
        
        \includegraphics[width=\linewidth]{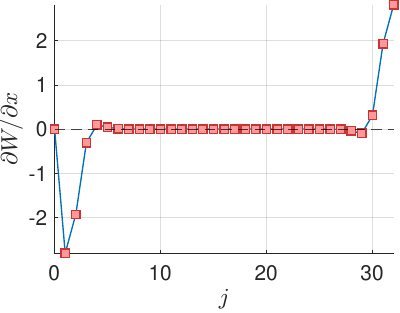}
       \caption{ circularly shifted kernel gradient $\frac{\partial W}{\partial x}_{Pad}$}

    \end{subfigure}

    \vspace{0.5em}  

    \caption{Zero padding of the SPH kernel gradient $\frac{\partial W}{\partial x}$, shown as the centre-row cross-section: (a) original kernel gradient; (b) circularly shifted, kernel wrapped to the corners for circular convolution via FFT.}
    \label{kernelPadding}
\end{figure}

For the work mentioned above, the linear convolution is computed, as shown in Equation (\ref{linearConv}), for an example of gradient $\frac{\partial f}{\partial \textbf{x}}$ at particle $i$:
\begin{equation}
    \frac{\partial f(\mathbf{x}_i)}{\partial \mathbf{x}} 
    = \left( f * \frac{\partial W}{\partial \mathbf{x}} \right)_i,
    \label{linearConv}
\end{equation}
where the kernel is evaluated at $\mathbf{x}_i - \mathbf{x}_j$ with no wrap-around. In this formulation, both the function $f$ (of length $m\times m$) and the kernel derivatives (of length $n\times n$) are zero-padded to length $(m+n-1)\times (m+n-1)$. However, under the Fourier Continuation framework, zero padding on $f$ will introduce a discontinuity which results in the Gibbs phenomenon. Therefore, in this work the circular convolution, as shown in Equation (\ref{circularConv}), is calculated to ensure that no padding is needed for $f$. 

\begin{equation}
    \frac{\partial f(\mathbf{x}_i)}{\partial \mathbf{x}} 
    = \sum_{j=1}^{N} f(\mathbf{x}_j) \frac{\partial W\!\left((\mathbf{x}_i - \mathbf{x}_j) \bmod L\right)}{\partial \mathbf{x}} \Delta x_j
    = \left( f \circledast \frac{\partial W}{\partial \mathbf{x}} \right)_i.
    \label{circularConv}
\end{equation}
Here $\circledast$ denotes the circular convolution operator, $L$ is the domain length, and $\bmod\, L$ maps the particle separation $\mathbf{x}_i - \mathbf{x}_j$ to its nearest periodic image in $[-L/2,\, L/2]^d$, ensuring the kernel is evaluated at the minimum-image distance across the periodic boundary. In this case, only the kernel derivatives need to be zero-padded to the size of the extended field $f_{\text{FC}}$. Figure \ref{kernelPadding} shows the original kernel derivative $\frac{\partial W}{\partial x}$ in the centre-row cross section (a) and its zero-padded counterpart (b), where the kernel values are wrapped to the corners of the extended grid in preparation for the FFT-based spectral SPH convolution.

   
         
        




\subsection{High-order Gaussian kernel functions}

\begin{figure}[h!]
            \centering
            \includegraphics[width=0.7\textwidth]{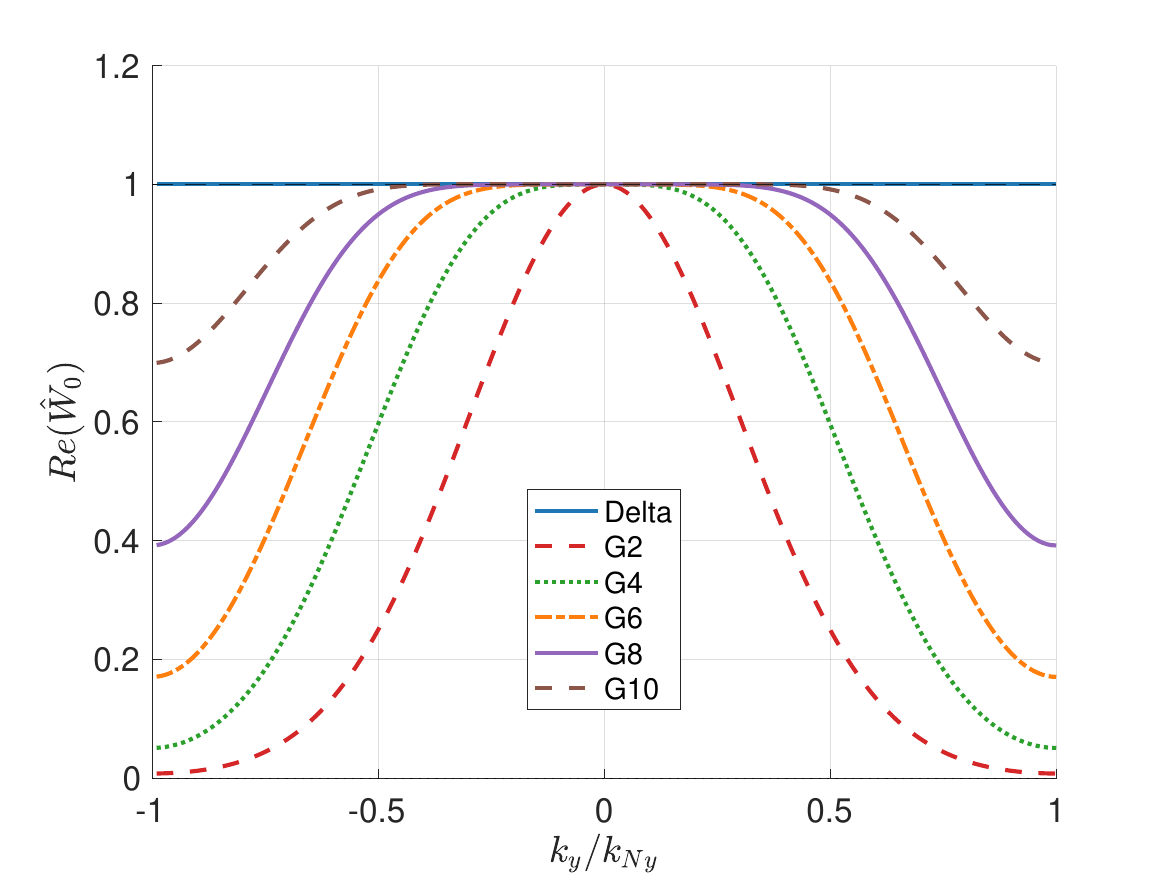}
            \caption{Real part of the Gaussian kernel  function $\text{Re}(\hat{W}_0)$ as a function of normalised wavenumber $k_y/k_{Ny}$, for kernel orders 
G2--G10. The Delta function is shown for reference.}
            \label{mimickDeltaFnction}
            \end{figure}

\begin{figure}[h!]
    \centering
   
    \begin{subfigure}{.48\textwidth}
        \centering
        \includegraphics[width=\linewidth]{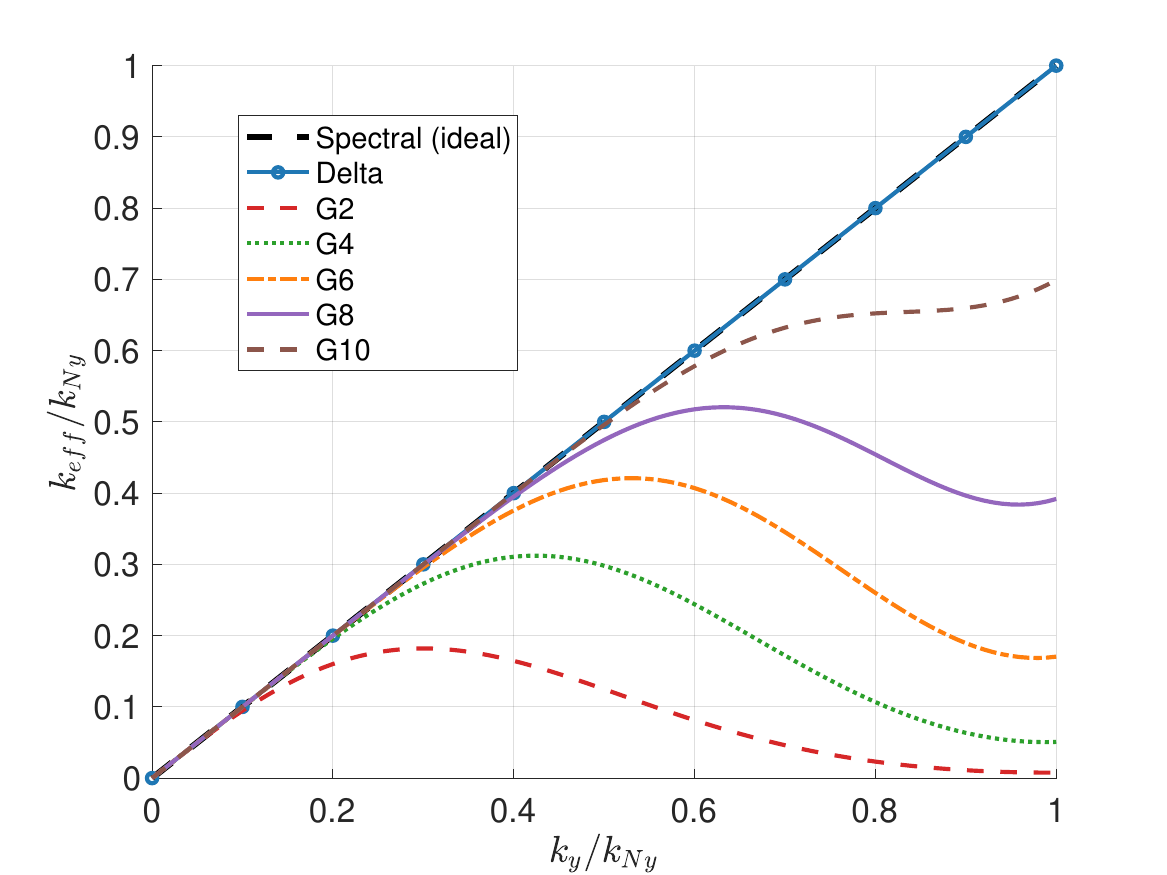}
        \caption{Gradient}
    \end{subfigure}
    \begin{subfigure}{.48\textwidth}
        \centering
        \includegraphics[width=\linewidth]{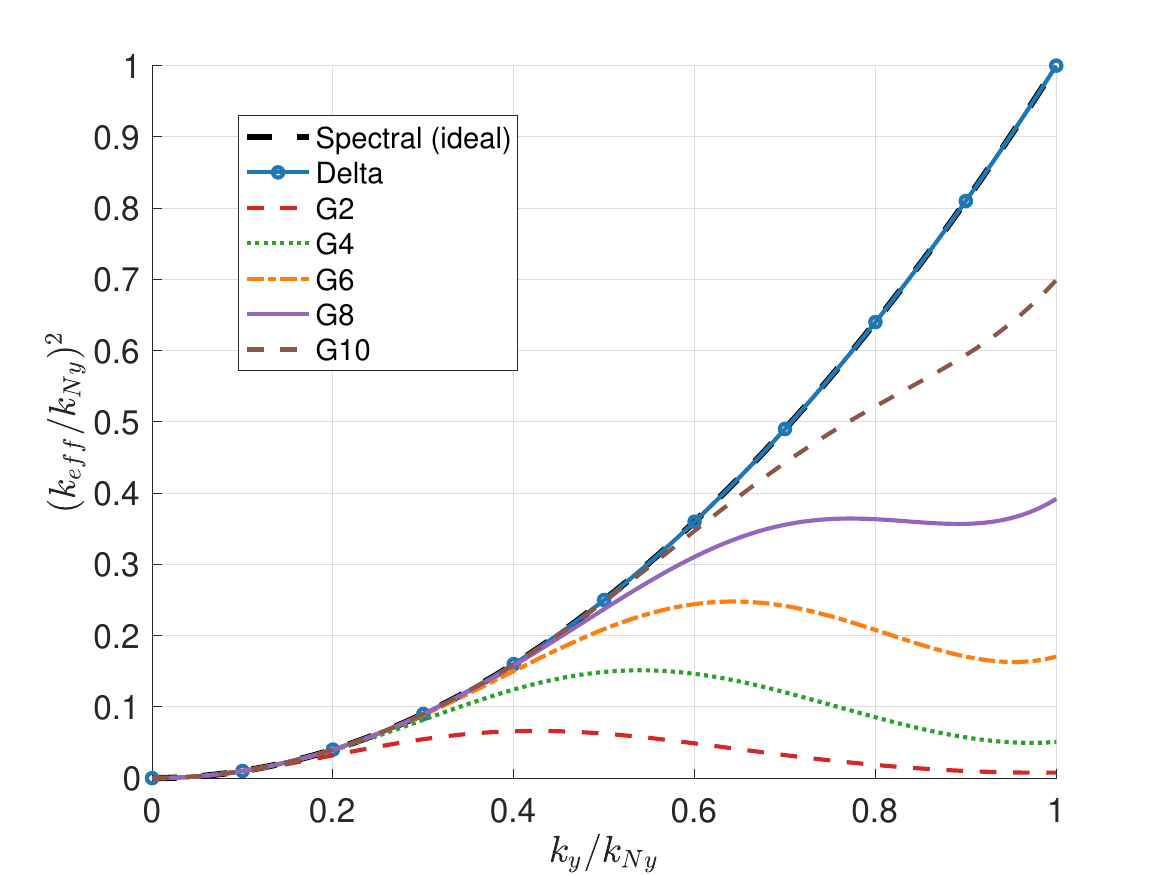}
        \caption{Laplaican}
    \end{subfigure}

    \vspace{0.5em}  

    \caption{Resolving power of different orders of the Gaussian kernel functions for the spectral SPH operators.}
    \label{resolvingPower}
\end{figure}

To maintain the high-order convergence rate, the Gaussian kernels proposed by Lind et al.~\cite{LIND2016JCP} are chosen as the SPH interpolation kernel. The $4th$ order and $6th$ order Gaussian kernel functions are implemented in this work and their formulation, and derivatives can be found in our previous work \cite{LINSpectralSPH}. Meanwhile, higher order kernel functions G8 and G10 are constructed as well to analyse the resolving power of different SPH kernels. Their formulations are given as follows:

\begin{align}
    W^{\text{G8}}(q) &= \alpha_{G8}\left(4 - 6q^2 + 2q^4 - \frac{q^6}{6}\right)e^{-q^2} \\
    W^{\text{G10}}(q) &= \alpha_{G10}\left(5 - 10q^2 + 4q^4 - \frac{2q^6}{6} + \frac{q^8}{24} \right)e^{-q^2}
\end{align}

\noindent where $q = r/h$ is the normalised distance, $h$ is the smoothing 
length and is set to $8dx$ in this work, and $\alpha_{G8}$, $\alpha_{G10}$ are normalisation constants chosen such that $\int W \, \text{d}\mathbf{x} = 1$. The detailed derivation can be found in the Appendix.
The order of the kernel function determines the accuracy of the SPH interpolation, furthermore, the closer the kernel function approximates the Dirac delta function, the closer the spectral SPH scheme is to the conventional spectral methods, as shown in Equation (\ref{kernelSpectral}). 

\begin{align}
    \frac{\partial f}{\partial \textbf{x}} &\approx f * \frac{\partial W}{\partial \textbf{x}} \nonumber\\
                                           &= \hat{f}\cdot{ik}\hat{\frac{\partial W}{\partial \textbf{x}}}
\label{kernelSpectral}
\end{align}
Here $\hat{\frac{\partial W}{\partial \textbf{x}}}$ is the Fourier transform of kernel gradient. Figure~\ref{mimickDeltaFnction} shows the real part of the Fourier 
transform of the high-order Gaussian kernels G2--G10 
as a function of the normalised wavenumber $k_y/k_{Ny}$ compared against the 
Dirac delta function $(\delta)$. As the kernel order increases, 
$\hat{W}_0$ approaches unity over a broader wavenumber range. This indicates that 
higher-order kernels resolve a greater portion of the frequency spectrum, which can also be observed from Figure \ref{resolvingPower}, where both the gradient and Laplacian operators 
of spectral SPH approximation with G8 and G10 kernel functions closely follow the ideal spectral curve over a broader wavenumber range compared to lower-order kernels. In the validation cases presented in Section~\ref{Testcasesvalidation}, 
G4 and G6 kernels are employed for low Reynolds number flows, while 
the higher-order G8 kernel is adopted for high Reynolds number flows to capture the fine-scale vortical structures and to keep the numerical stability.


\section{Fourier Continuation (FC) Algorithm}
\label{sec:fc:1d}
This section introduces the Fourier Continuation (FC) extension algorithm for extending the non-periodic computational domain to be periodic.
Consider a function $f$ sampled on a uniform grid
$y_j = j\,\Delta y$, $j = 0, 1, \ldots, N-1$, with spacing
$\Delta y = 1/N$. The FC procedure constructs a smooth continuation
$f_{\mathrm{cont}}$ on the extension region
$[1,\; 1 + d\,\Delta y]$, where $d$ is the number of extension points,
such that the extended function
\begin{equation}
    F(y) =
    \begin{cases}
        f(y),              & y \in [0, 1],              \\
        f_{\mathrm{cont}}(y), & y \in (1,\; 1 + d\,\Delta y],
    \end{cases}
    \label{eq:fc_extended}
\end{equation}
is both smooth up to $C^p$ and periodic on $[0,\; 1 + d\,\Delta y]$.


The FC scheme proposed in this work can be achieved in the following three steps: boundary polynomial fitting, extrapolation and blending.
 
\subsection{Step 1: Boundary Polynomial Fitting}
\label{sec:fc:polyfit}
 
Let $C$ denote the number of grid points selected near each boundary for polynomial fitting, as shown in Figure \ref{fig:boundaryExtraction}. To express these points in a local coordinate system centred at each wall, we define the wall-local coordinates
\begin{equation}
    \xi^{r}_{i} = y_{N-1-C+i} - 1 \in [-C\Delta y,\; 0],
    \qquad
    \xi^{\ell}_{i} = y_{i} \in [0,\; C\Delta y],
    \qquad i = 0, \ldots, C-1,
    \label{eq:shifted_coords}
\end{equation}
where the origin $\xi = 0$ coincides with the respective wall in each case. Here $\xi^{r}_i \leq 0$ measures distance inward from the right wall ($y = 1$), and $\xi^{\ell}_i \geq 0$ measures distance inward from the left wall ($y = 0$). Expressing the boundary points in these local coordinates improves the conditioning of the polynomial fit, since all $C$ points lie within a small neighbourhood of the origin regardless of their global position.
A degree-$p$ polynomial is fitted to the $C$ boundary values near both boundaries by solving the overdetermined linear system
\begin{equation}
    \mathbf{V}\,\mathbf{c} = \mathbf{f},
    \label{eq:vandermonde}
\end{equation}
by using the least-squares condition, where
\begin{equation}
    \mathbf{c} = \arg\min_{\mathbf{c} \in \mathbb{R}^{p+1}}
    \left\| \mathbf{V}\mathbf{c} - \mathbf{f} \right\|^2.
    \label{eq:least_squares}
\end{equation}
Here, the solution $\mathbf{c} = [c_p, c_{p-1}, \ldots, c_1, c_0]^T$ 
are the coefficients of the fitted polynomial
\begin{equation}
    p(\xi) = c_p \xi^p + c_{p-1} \xi^{p-1} + \cdots + c_1 \xi + c_0,
\end{equation}
$\mathbf{f} = [f_0, f_1, \ldots, f_{C-1}]^T$ is the vector of function
values at the $C$ boundary points, and $\mathbf{V} \in \mathbb{R}^{C
\times (p+1)}$ is the Vandermonde matrix
\begin{equation}
    \mathbf{V} =
    \begin{bmatrix}
        \xi_0^p     & \xi_0^{p-1}     & \cdots & \xi_0   & 1 \\
        \xi_1^p     & \xi_1^{p-1}     & \cdots & \xi_1   & 1 \\
        \vdots      & \vdots          & \ddots & \vdots  & \vdots \\
        \xi_{C-1}^p & \xi_{C-1}^{p-1} & \cdots & \xi_{C-1} & 1
    \end{bmatrix}.
    \label{eq:vandermonde_matrix}
\end{equation}
Since $C > p + 1$, the linear system~\eqref{eq:vandermonde} is overdetermined
and the least-squares solution is obtained via QR decomposition:
\begin{equation}
    \mathbf{V} = \mathbf{Q}\mathbf{R},
    \qquad
    \mathbf{c} = \mathbf{R}^{-1}\mathbf{Q}^T\mathbf{f}.
    \label{eq:qr}
\end{equation}
This yields the fitted polynomials
\begin{align}
    p_r(\xi) &= \sum_{j=0}^{p} c_j\,\xi^{p-j},
    \qquad \xi = y - 1 \leq 0,
    \label{eq:p_right} \\
    p_\ell(\xi) &= \sum_{j=0}^{p} d_j\,\xi^{p-j},
    \qquad \xi = y \geq 0,
    \label{eq:p_left}
\end{align}
where $\mathbf{c} = [c_0,\ldots,c_p]^T$ and
$\mathbf{d} = [d_0,\ldots,d_p]^T$ are the coefficient vectors for the
right and left boundaries respectively, and $j$ is the summation index 
with exponent $p-j$ decreasing from $p$ to $0$.
\subsection{Step 2: Polynomial Extrapolation}
\label{sec:fc:extrapolation}
 
The extension points are located at
\begin{equation}
    y_k = 1 + k\,\Delta y, \qquad k = 1, 2, \ldots, d.
    \label{eq:gap_points}
\end{equation}
Both fitted polynomials are extrapolated into the extension area by evaluating them
at coordinates outside their fitting regions. Specifically:
 
\begin{itemize}
    \item $p_r$ is extrapolated forward past $y = 1$ by evaluating
    at $\xi = k\,\Delta y > 0$:
    \begin{equation}
        f_r(y_k) = p_r(k\,\Delta y)
        = \sum_{j=0}^{p} c_j\,(k\,\Delta y)^{p-j};
        \label{eq:f_right_gap}
    \end{equation}
 
    \item $p_\ell$ is extrapolated backward past $y = 0$ by
    evaluating at $\xi = -(d-k)\,\Delta y \leq 0$, and the resulting
    values are placed into the extension area:
    \begin{equation}
        f_\ell(y_k) = p_\ell\!\left(-(d-k)\,\Delta y\right)
        = \sum_{j=0}^{p} d_j\,\bigl(-(d-k)\,\Delta y\bigr)^{p-j}.
        \label{eq:f_left_gap}
    \end{equation}
\end{itemize}

The results of boundary points extraction, polynomial fitting and extrapolation are presented in the following Figure \ref{FCStep2_3}.


\begin{figure}[h!]
    \centering
    \begin{subfigure}{.48\textwidth}
        \centering
        \includegraphics[width=\linewidth]{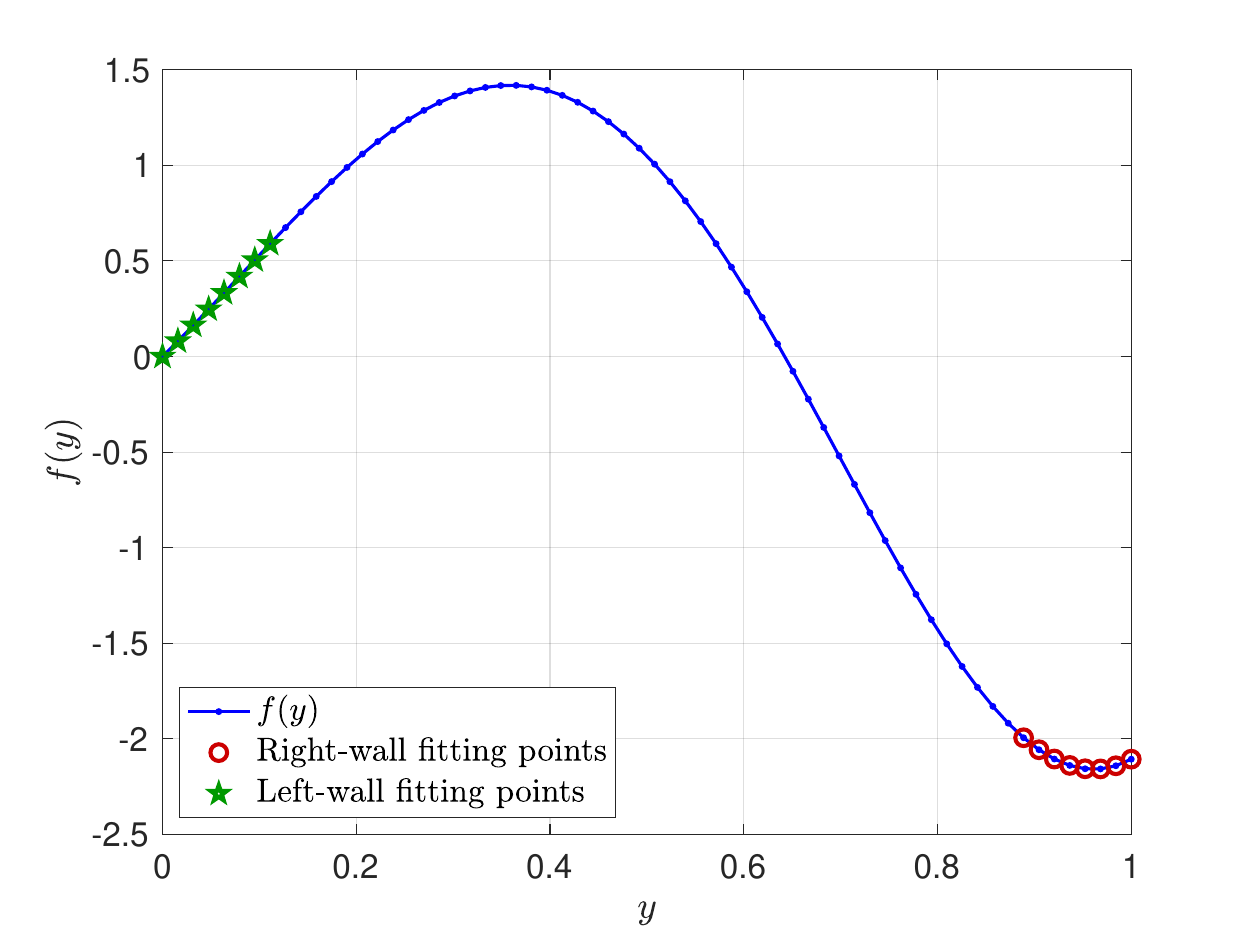}
        \caption{Boundary points extraction}
        \label{fig:boundaryExtraction}
    \end{subfigure}
    \begin{subfigure}{.48\textwidth}
        \centering
        \includegraphics[width=\linewidth]{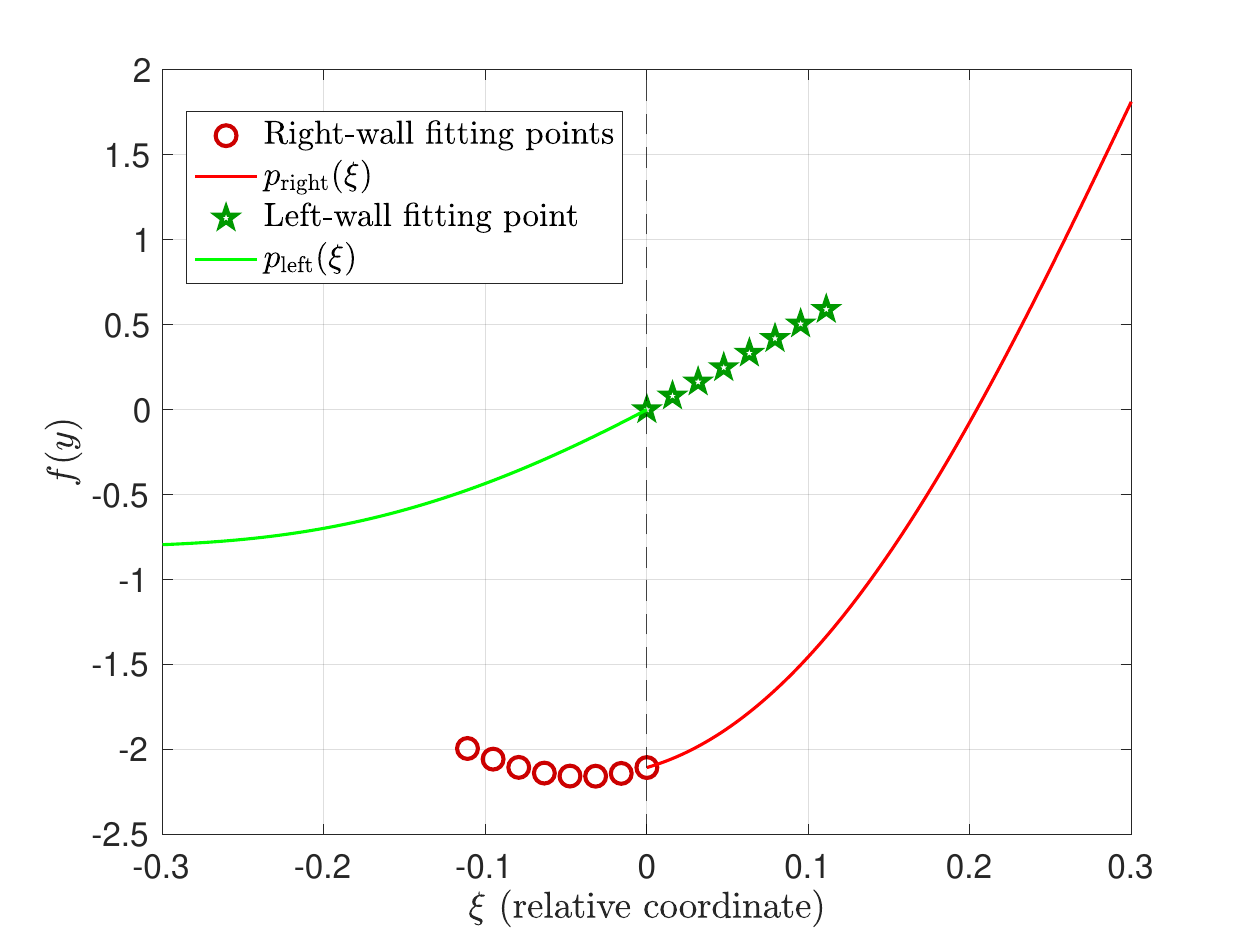}
        \caption{Polynomial fitting and extrapolation}
        \label{fig:polynomialFitting}
    \end{subfigure}
    \vspace{0.5em}
    \caption{Polynomial fitting and extrapolation of the Fourier continuation on the example function $f = e^y \sin(5y) + 0.5y^3.$}
    \label{FCStep2_3}
\end{figure}

 
\subsection{Step 3: Smooth Blending}
\label{sec:fc:blending}
 
\begin{figure}[h!]
    \centering
   
    \begin{subfigure}{.48\textwidth}
        \centering
        \includegraphics[width=\linewidth]{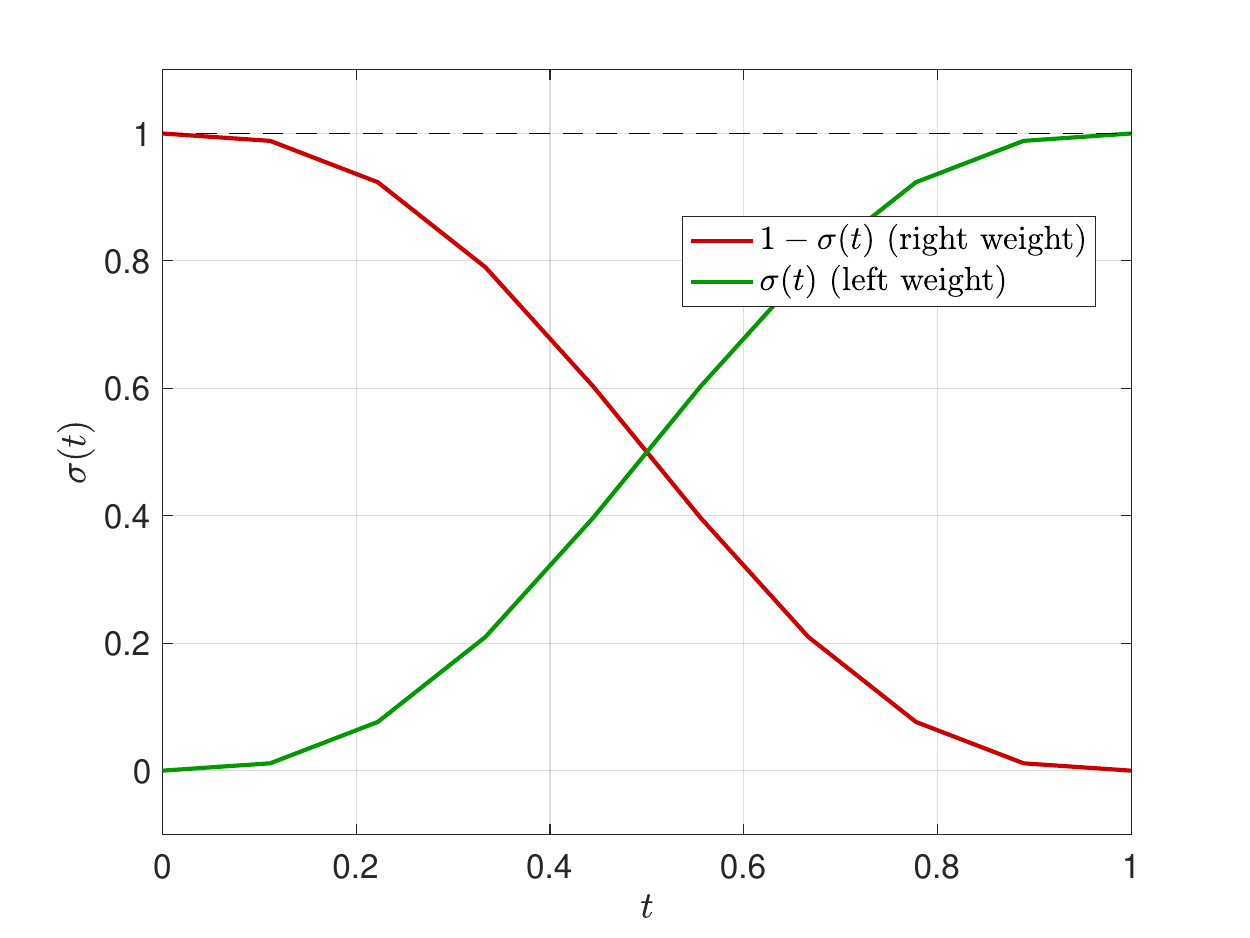}
        \caption{Blending function}
        \label{blendingFunction}
    \end{subfigure}
    \begin{subfigure}{.48\textwidth}
        \centering
        \includegraphics[width=\linewidth]{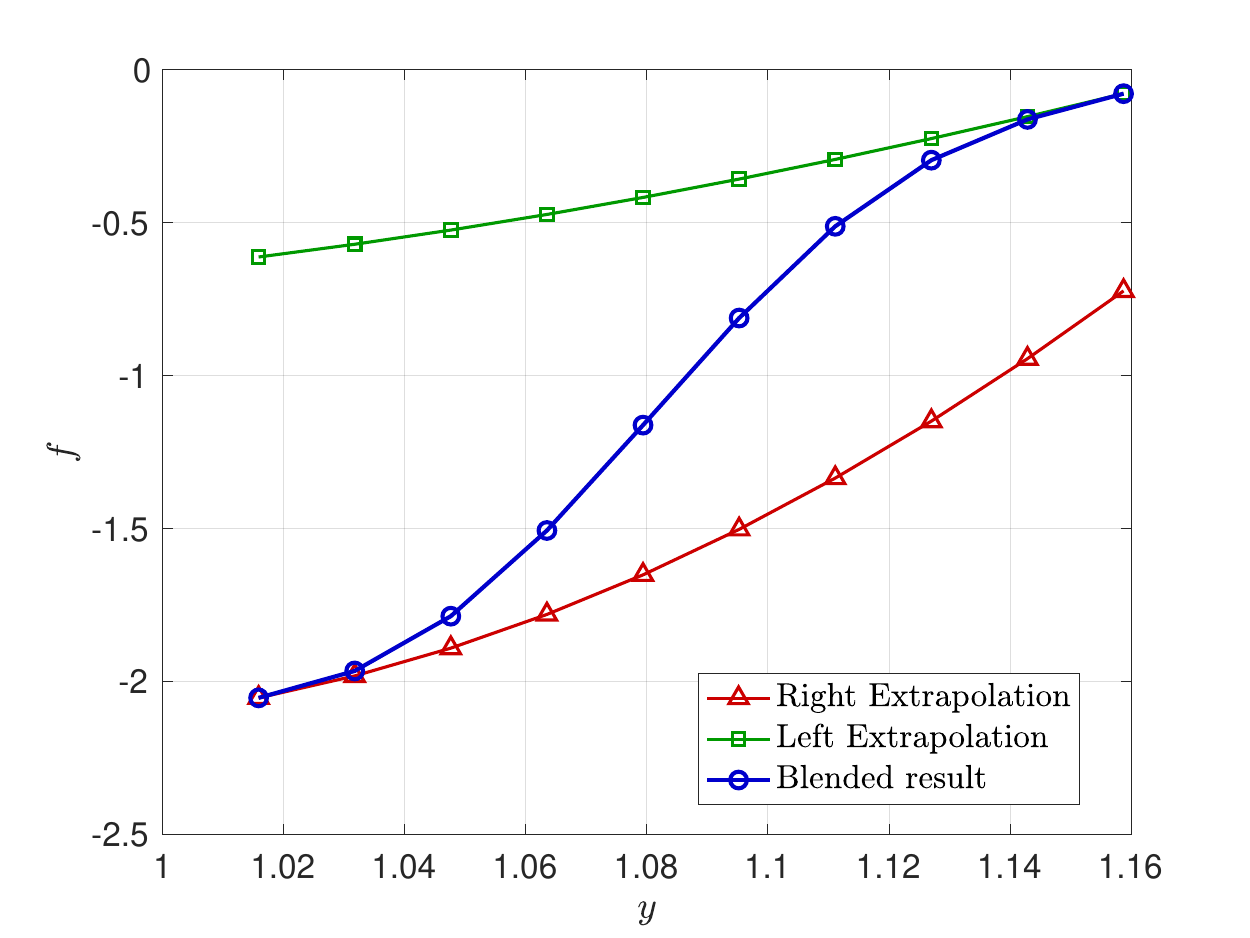}
        \caption{FC-extended area}
        \label{FC-extended area}
    \end{subfigure}

    \vspace{0.5em}  

    \caption{Blending function $\sigma(t)$ and FC-extended area of the example function.}
    \label{FCStep_blending}   
\end{figure}

After the extrapolation,  $f_r$ and $f_\ell$ are blended to be continuously linked within the extension area
using a smooth $S^p$ blending function $\sigma(t)$:
\begin{equation}
    f_{\mathrm{cont}}(y_k)
    = \bigl(1 - \sigma(t_k)\bigr)\,f_r(y_k)
    + \sigma(t_k)\,f_\ell(y_k),
    \qquad t_k = \frac{k-1}{d-1} \in [0,1],
    \label{eq:fc_blend}
\end{equation}
where $\sigma$ is a $5th$ order polynomial shown in Figure \ref{blendingFunction} :
\begin{equation}
    \sigma(t) = 10t^3 - 15t^4 + 6t^5,
    \label{eq:sigma}
\end{equation}
satisfying
\begin{equation}
    \sigma(0) = 0, \quad \sigma(1) = 1, \quad \sigma'(0) = 0,
    \quad \sigma'(1) = 0 
    \label{eq:sigma_bc}
\end{equation}
Conditions~\eqref{eq:sigma_bc} ensure that $\sigma(t)$ transitions 
smoothly from 0 to 1 across the discontinuity, so that $f_{\mathrm{cont}}$ 
connects smoothly to $f(1)$ at the start of the periodic extension interval and to $f(0)$ at 
the end of it, as shown in Figure \ref{FC-extended area}. This makes the extended function $F(y)$ periodic and 
smooth on $[0,\; 1 + d\,\Delta y]$, which is required for accurate FFT-based spectral SPH discretisation.

All the three steps are illustrated in the following Figure \ref{FCIllu}, in which a non-periodic function $f(y)=e^ysin(5y)+0.5y^3$ with $y\in[0,1]$ is extended beyond its right boundary $y=1$.

\begin{figure}[h!]
            \centering
            \includegraphics[width=1.0\textwidth]{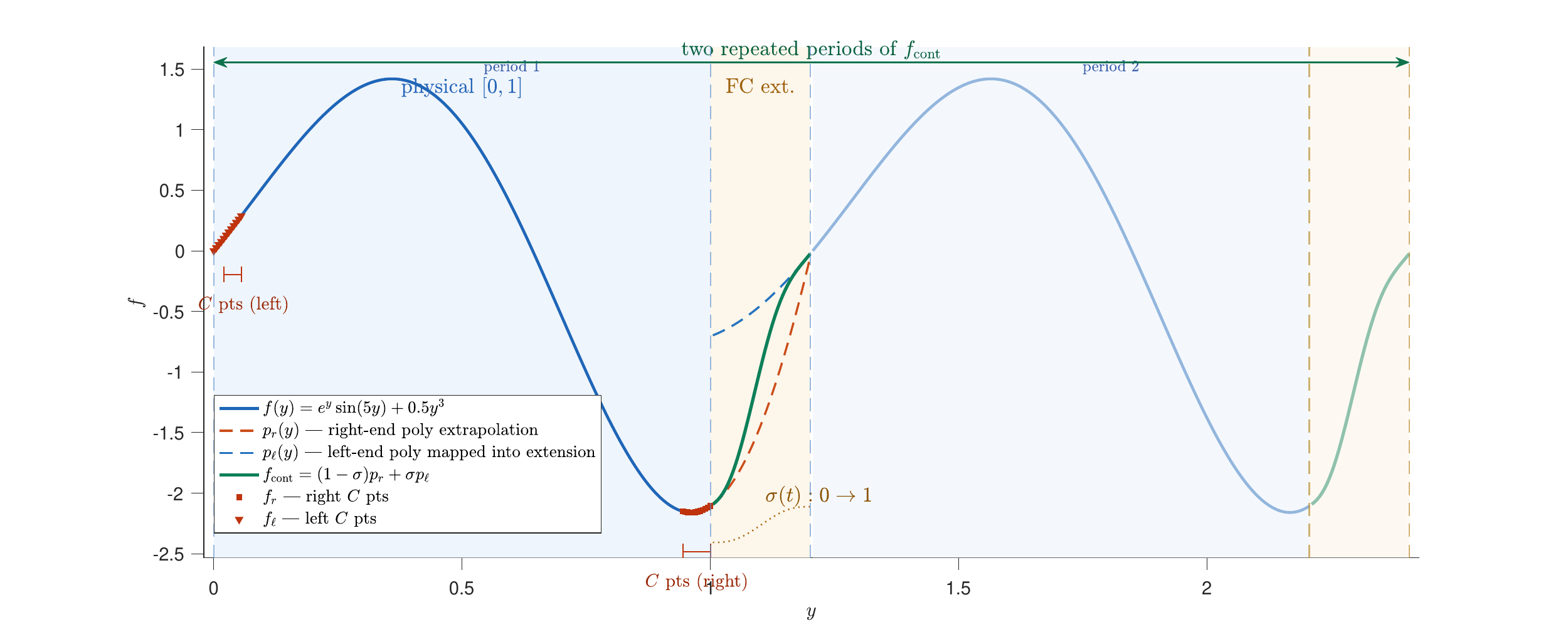}
            \caption{Fourier continuation of a non-periodic function $f(y)=e^ysin(5y)+0.5y^3$ on the interval of $[0,1]$.}
            \label{FCIllu}
            \end{figure}

\subsection{Incorporation of the Neumann Boundary Condition}
For the wall-bounded flows simulated in this work, pressure satisfies Neumann boundary conditions, therefore modifications are needed for the aforementioned FC algorithm to account for the pressure gradient at the boundaries $\frac{\partial P}{\partial \textbf{n}}=0$.  The derivative of the fitted polynomial $p(\xi)$ is
\begin{equation}
    p'(\xi) = pc_0\xi^{p-1} + (p-1)c_1\xi^{p-2} + \cdots + c_{p-1},
\end{equation}
and at $\xi = 0$, all terms vanish except the last, which gives
\begin{equation}
    c_{p-1} = 0.
\end{equation}
Therefore, enforcing $\frac{\partial P}{\partial \mathbf{n}}=0$ 
reduces to the single condition $c_{p-1} = 0$. This can be easily achieved by adding one row in the Vandermonde matrix:

\begin{equation}\label{FCNeumann}
\mathbf{V}_{\text{Neu}} = \left[\begin{array}{c}
\mathbf{V} \\
\hline
0 \quad 0 \quad \cdots \quad 1 \quad 0
\end{array}\right], \qquad
\mathbf{b} = \begin{bmatrix} \mathbf{f} \\ 0 \end{bmatrix}
\end{equation}
to enforce:
\begin{equation}\label{}
P'(0)=c_{p-1}=\frac{\partial P}{\partial \textbf{n}}|_{\xi=0}=0.
\end{equation}

\subsection{Two-Dimensional FC Extension for Wall-Bounded Domains}
\label{sec:fc:2d}
 
For the wall-bounded flows, the two spatial directions are treated
differently:
 
\begin{itemize}
    \item \textbf{Streamwise direction} $x$: the flow is inherently
    periodic in the $x$ direction, so no FC extension is required. The field is simply wrapped
    periodically and implicitly by the discrete Fourier transform (DFT).

    \item \textbf{Wall-normal direction} $y$: the Dirichlet
    conditions require FC. The one-dimensional FC
    extension explained above is applied:
   \begin{equation}
    F(x,\, y) =
    \begin{cases}
        f(x,\, y),          & y \in [y_0,\, y_{N-1}],   \\
        f_{\mathrm{cont}}(x,\, y),    & y \in [y_N,\, y_{N+d-1}],
    \end{cases}
    \label{eq:fc_2d}
\end{equation}
where $f_{\mathrm{cont}}(\,\cdot\,,\, y)$ denotes the FC continuation
constructed at streamwise location $y$.
\end{itemize} 

The resulting extended field $F$ is defined on the
enlarged grid of size $(N + d) \times (N)$ and is periodic in both
directions. This enables the FFT-based spectral SPH scheme without the Gibbs oscillation.



While the proposed FC scheme is based on a $5$th order polynomial fitting and extrapolation, the well-established FC(Gram) approach of Albin $\&$ Bruno~\cite{Albin2011FCNavierStokes} constructs the periodic extension via projection onto a Gram polynomial basis, where the extension matrices are precomputed once and applied at runtime as matrix-vector multiplications. This is computationally cheaper and more stable. The implementation of this FC variant is reserved for future work and the readers can refer to \cite{FC2020SPECTRA} for an open source pseudospectral solver SPECTER with the FC(Gram) implementation, and to Lyon \cite{Lyon2011FastFC} for a fast FC algorithm.

\section{A FC-based spectral ISPH framework} 
\label{wholeframework}

\subsection{Spatial discretisation of pressure}
 Similar to our previous work in \cite{LINSpectralSPH}, the spectral PPE solver is implemented to reduce the cost and avoid the linear iteration solver. For pressure with Neumann boundary conditions, the PPE can be efficiently solved using discrete cosine transforms (DCT) \cite{FUKApoisFFT}. For collocated Neumann boundary conditions, the first type of discrete cosine transform (DCT-I) in two dimensions is given by
\begin{equation}\label{DCTI4PPE}
\begin{split}
\hat{P}_{kl} = P_{0,0} + (-1)^k P_{N-1,0} + (-1)^l P_{0,M-1} + (-1)^{(k+l)} P_{N-1,M-1} \\
+ 2\sum^{N-2}_{i=1}\sum^{M-2}_{j=1} P_{ij}
\cos\!\left(\frac{\pi i k}{N-1}\right)
\cos\!\left(\frac{\pi j l}{M-1}\right).
\end{split}
\end{equation}
For wall-bounded flows where the pressure is periodic in $x$ and subject to Neumann boundary conditions in $y$, the PPE is solved using a DFT in $x$ and a DCT-I in $y$:
\begin{equation}\label{DCTDFT_PPE}
\begin{split}
\hat{P}_{kl} = \sum^{N-1}_{m=0} \left[P_{m,0} + (-1)^l P_{m,M-1}\right] e^{\left(-\frac{2\pi m k \mathrm{i}}{N}\right)} \\
+ 2\sum^{N-1}_{m=0}\sum^{M-2}_{j=1} P_{mj}\,
\cos\!\left(\frac{\pi j l}{M-1}\right)
e^{\left(-\frac{2\pi m k \mathrm{i}}{N}\right)}.
\end{split}
\end{equation}
Applying the discrete transform to both sides of the pressure Poisson equation diagonalises the Laplacian operator in spectral space. In Fourier space, the PPE reduces to a simple algebraic operation:
\begin{equation}\label{division4PPE}
\hat{\phi}_{kl}=\frac{\hat{b}_{kl}}{\lambda_k^{2}+\lambda_l^{2}},
\end{equation}
where $\hat{b}_{kl}$ is the transformed right-hand side of the PPE in Equation~(\ref{PPE}), and $\lambda_k$, $\lambda_l$ are the eigenvalues of the discrete Laplacian in the $x$ and $y$ directions, respectively. For the wall-bounded configuration, $\lambda_k$ corresponds to the DFT modes and $\lambda_l$ to the DCT-I modes. Details  of corresponding transformation types and equations of wavenumber for a broader range of boundary conditions can be found in \cite{FUKApoisFFT}. 




\subsection{Integration of Fourier continuation into the spectral ISPH solver}
\begin{figure}[h!]
            \centering
            \includegraphics[width=1.0\textwidth]{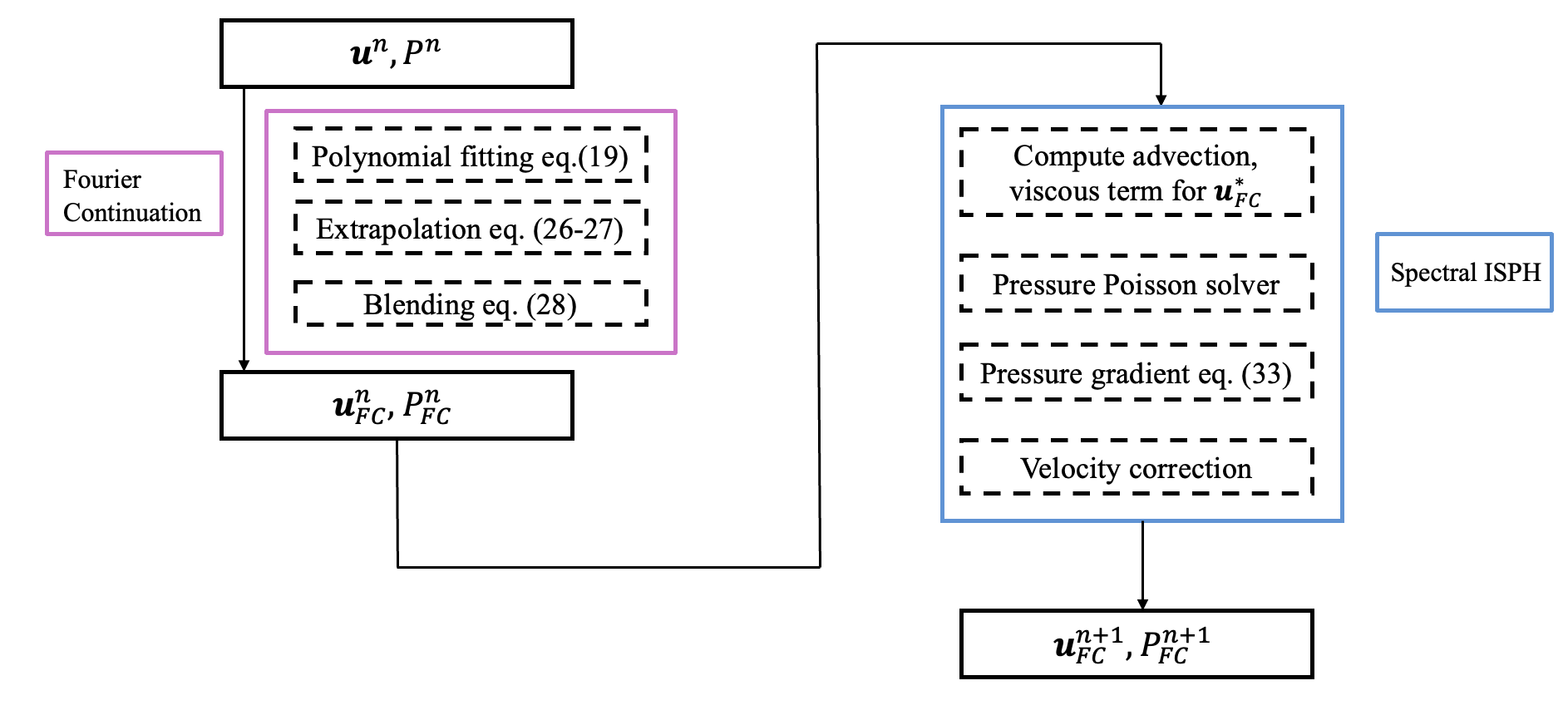}
            \caption{The computational procedure of one time step by the Fourier continuation-based spectral ISPH scheme.}
            \label{FCFlowChart}
            \end{figure}

The Fourier continuation algorithm introduced above can be served as a pre--processing module of the incompressible spectral ISPH solver. The procedure for solving one time step is shown in Figure \ref{FCFlowChart}. In the current solver, the pressure is obtained through a spectral Neumann 
PPE solver on the physical fluid domain. Alternatively, the PPE could be 
solved on the FC-extended periodic domain using a fully periodic solver, 
as demonstrated in~\cite{FC2020SPECTRA}. We reserve this for future development.

\section{Test cases validation} \label{Testcasesvalidation}

\subsection{Convergence tests} \label{convergenceOperator}
\begin{figure}[h!]
    \centering
   
    \begin{subfigure}{.48\textwidth}
        \centering
        \includegraphics[width=\linewidth]{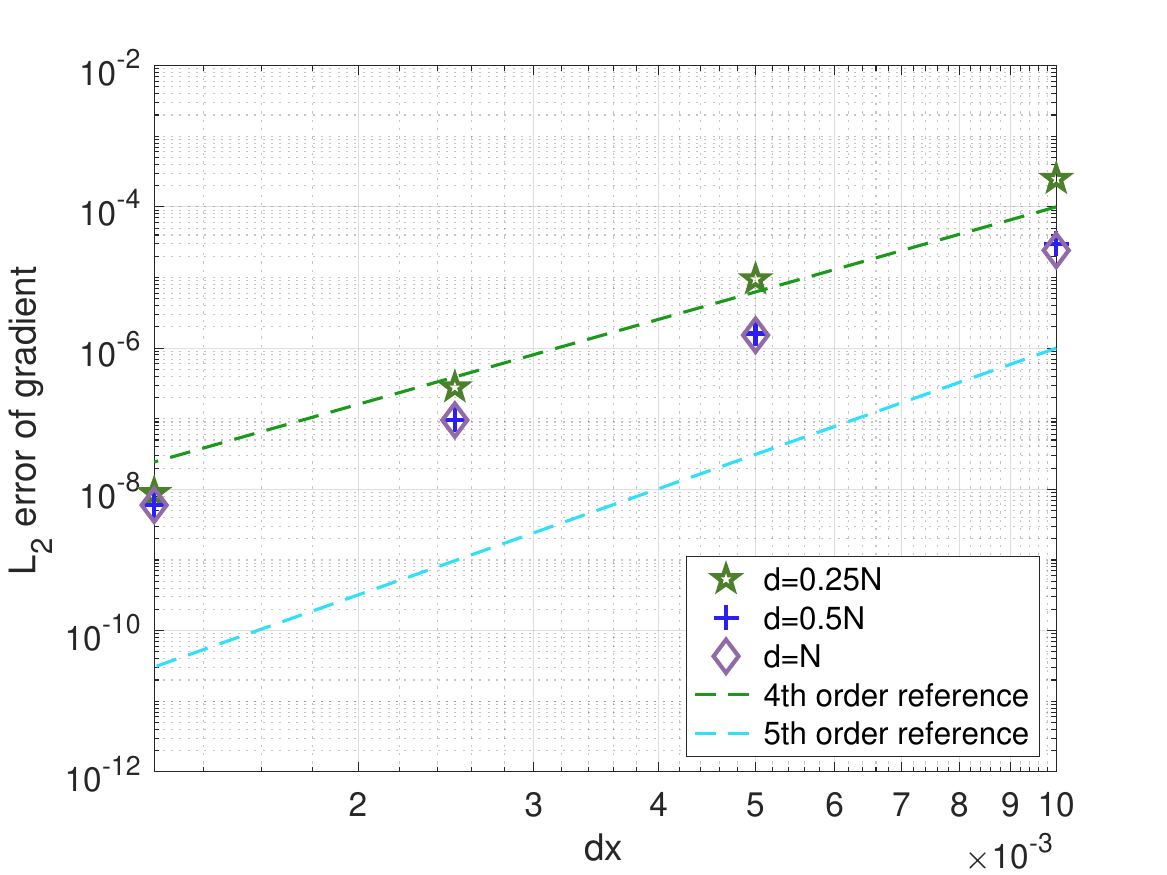}
        \caption{gradient}
    \end{subfigure}
    \begin{subfigure}{.48\textwidth}
        \centering
        \includegraphics[width=\linewidth]{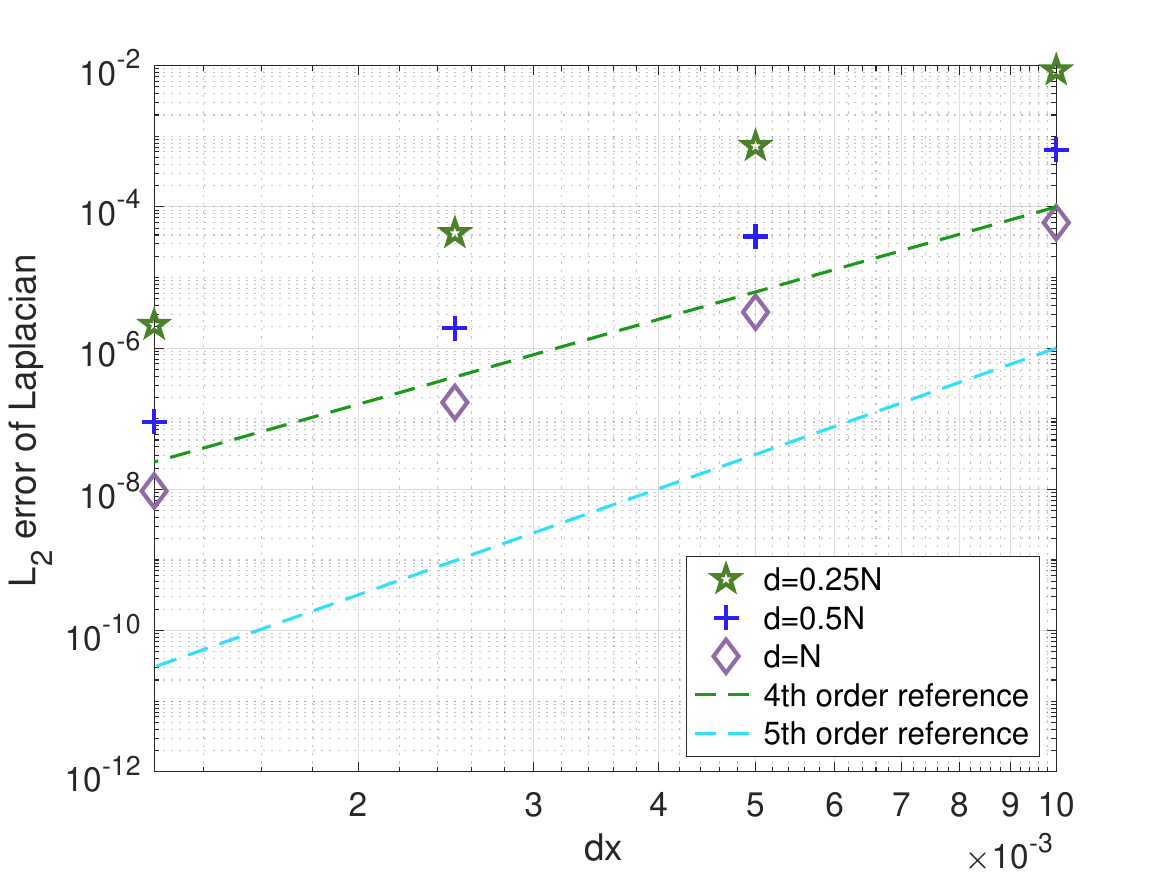}
        \caption{Laplacian}
    \end{subfigure}

    \vspace{0.5em}  

    \caption{Convergence rate of gradient and Laplacian operators by the FC-based spectral SPH scheme under different extension length.}
    \label{change_d_ext}
\end{figure}

A convergence study is conducted to assess the accuracy of the FC-based spectral SPH operators and the influence of key FC parameters, namely the extension length and the fitted polynomial degree. Here the convergence rate is plotted to indicate the convergence rate. Throughout this paper, the $L_2$ norm of a function $f$ with size $n_x\times n_y$ in 2D is defined as:

\begin{equation}\label{L_2 norm}
L_2(f)=\sqrt{\frac{1}{n_x}\frac{1}{n_y}\sum^{n_x}_{i=1}\sum^{n_y}_{j=1}(f_{SPH}(i,j)-f_{analytical}(i,j))^2} .
     \end{equation}
The test function is chosen as:

\begin{equation}\label{testFun10}
f(x,y)=sin(2\pi x)y(1-y)+0.5cos(4\pi x)y^2(1-y)^2,
     \end{equation}
which satisfies no-slip boundary condition in $y$ direction, and periodic boundary condition in $x$ direction. Figure \ref{change_d_ext} shows the convergence behaviour of the gradient and Laplacian operators by the FC-based spectral SPH scheme, with different Fourier continuation extension length $d$, varying from a quarter of the original domain $d=0.25N$ to a full domain $d=N$. From the convergence plot, both the gradient and the Laplacian operators achieve more than $4$th order convergence rate. These are consistent with the degree $5$ of the FC fitting polynomial. Meanwhile, increasing the extension length improves accuracy and reduces the $L_2$ error of the derivative operators. Using $d=0.25N$ and $d=N$, the $L_2$ errors are of the order $10^{-4}$ and $10^{-7}$ respectively. However, the convergence rate remains unchanged, and the larger extension increases the computational cost from $1.25N \times N$ to $2N \times N$.
Therefore, an extension length for FC of $d=0.25N$ is adopted for all simulations presented in this work.

\begin{figure}[h!]
            \centering
            \includegraphics[width=0.7\textwidth]{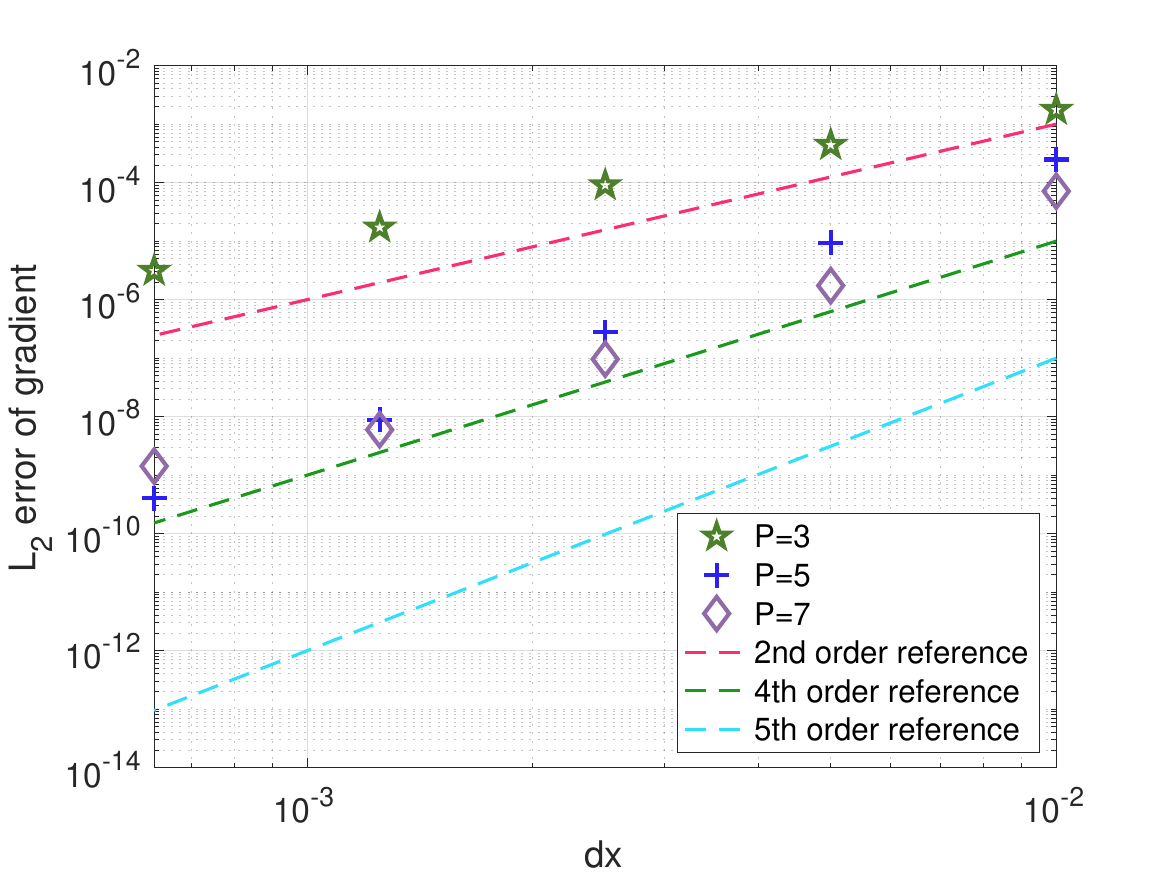}
            \caption{Convergence rate of gradient operator by the FC-based spectral SPH scheme under different fitting polynomial orders.}
            \label{changeP_order}
            \end{figure}

The influence of fitting polynomial order with the change of the $L_2$ norm is shown in Figure \ref{changeP_order}. Since implementing a $P$th order fitting polynomial will result in a $(P+1)$th order of Fourier continuation, the gradient operator achieves $\mathcal{O}(h^P)$
 convergence after differentiation. However, this is limited by the the 4th-order consistency of the G4 SPH kernel, which can be observed from the $4$th-order convergence gradient with $P=5$ and $P=7$.
Furthermore, using a higher polynomial order increases the cost and complexity of the FC extension, and increasing 
$P$ beyond yields no further accuracy gains due to the G4 SPH kernel function. Therefore, $P=5$ is adopted for the Fourier continuation scheme proposed in this work, as it provides a sufficient balance between accuracy and computational efficiency.

All the convergence results discussed validate that the FC-based spectral SPH scheme can achieve the high-order accuracy for non-periodic computational domain with an acceptable extra computational cost which is $25\%$ of the original domain.

\subsection{Gaussian vortex convection}
The previous section discussed the convergence behaviour of the derivative operators for a prescribed test function. In this section, a Gaussian vortex convection problem is solved to validate the proposed scheme for the following linear advection equation:

\begin{equation}\label{linearAdvection}
\frac{\partial u}{\partial t}+C_x\frac{\partial u}{\partial x}+C_y\frac{\partial u}{\partial x}=0.
     \end{equation}
The initial condition is defined as $u_0=e^{\alpha r^2}$, with $r=\sqrt{(x-x_0)^2+(y-y_0)^2}$ and the initial vortex centre $(x_0,y_0)=(0.5,0.5)$ lies in the centre of the computation domain. The vortex is convected with the speed $C_x=1$ m/s and $C_y=0.5$ m/s, which gives the following analytical solution of $u$:

\begin{equation}\label{GaussianVortexAnalytical}
        u=e^{-\alpha(x-(x_0+C_xt))^2+(y-(y_0+C_yt))^2}.
     \end{equation}

\begin{figure}[h!]
    \centering
    \begin{subfigure}{.32\textwidth}
        \centering
        \includegraphics[width=\linewidth]{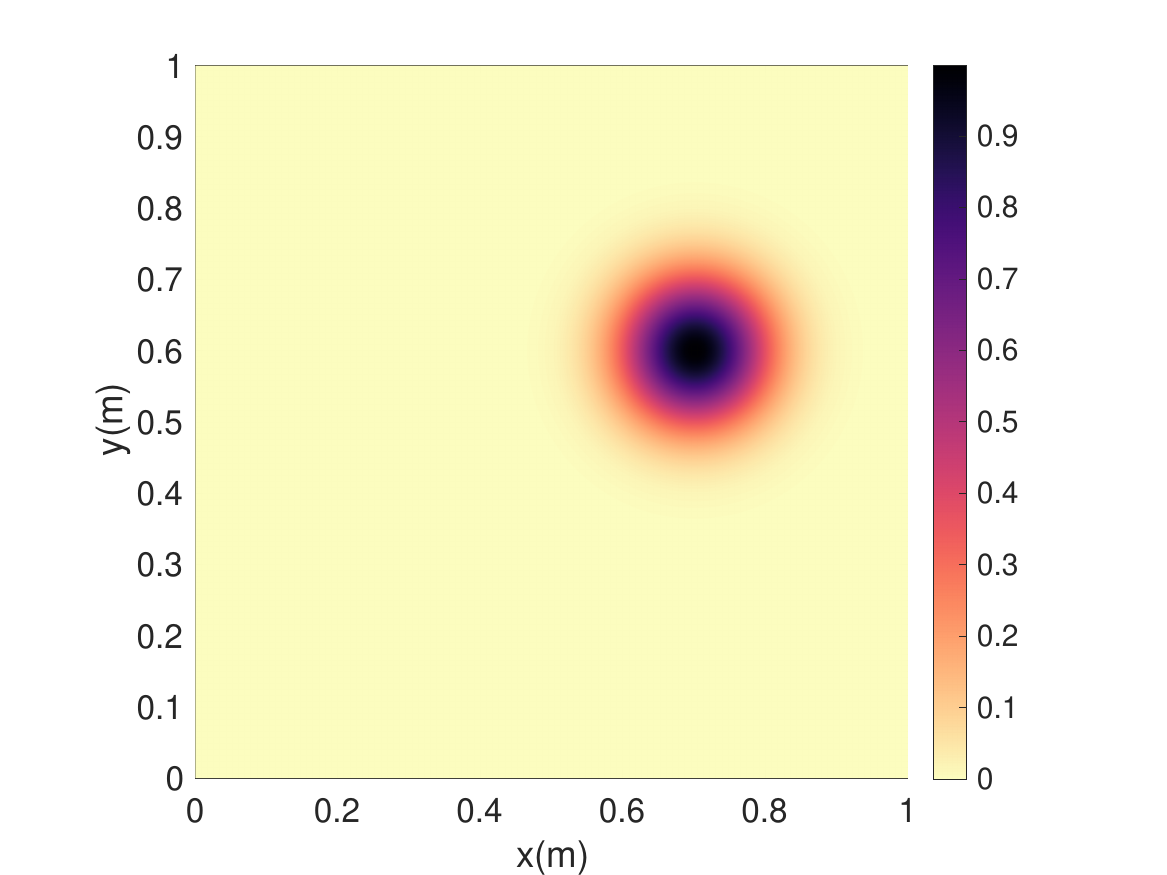}
        \caption{t=0.2s}
    \end{subfigure}
    \begin{subfigure}{.32\textwidth}
        \centering
        \includegraphics[width=\linewidth]{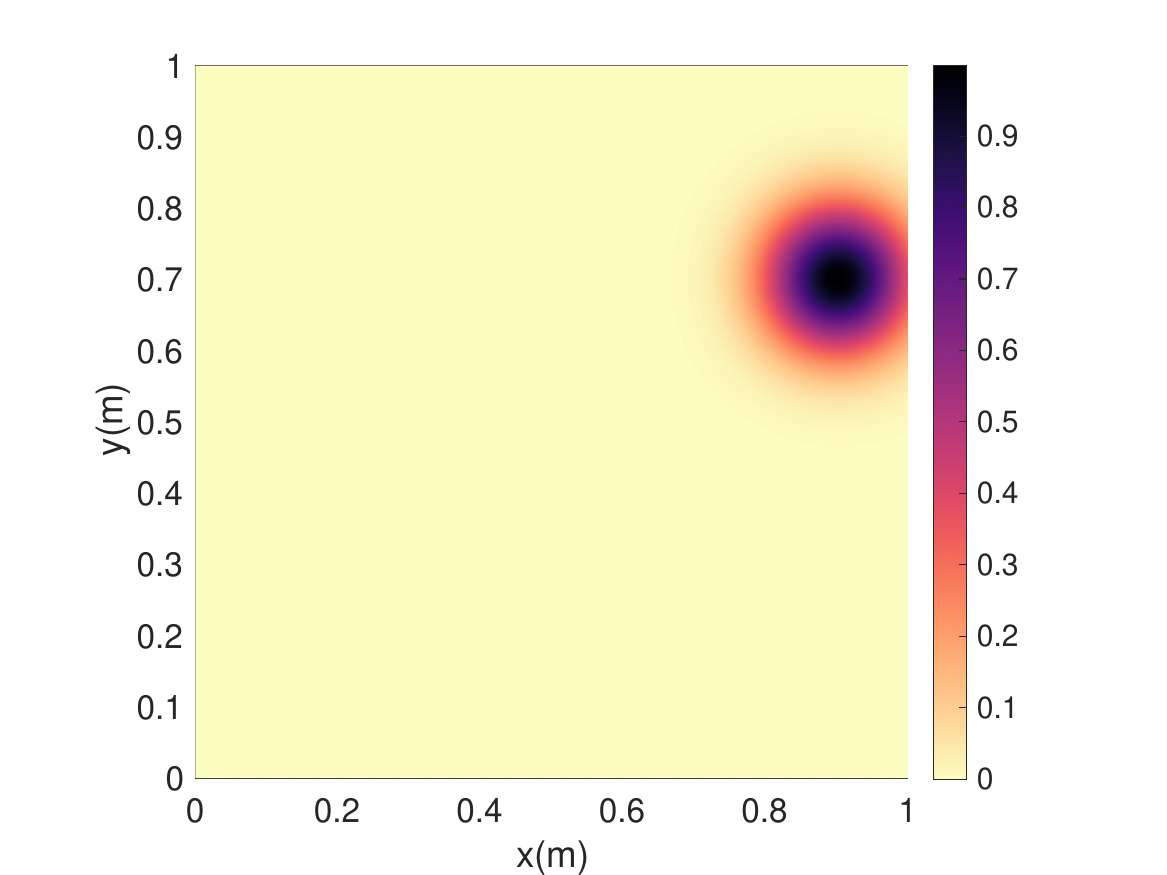}
        \caption{t=0.4s}
    \end{subfigure}
    \begin{subfigure}{.32\textwidth}
        \centering
        \includegraphics[width=\linewidth]{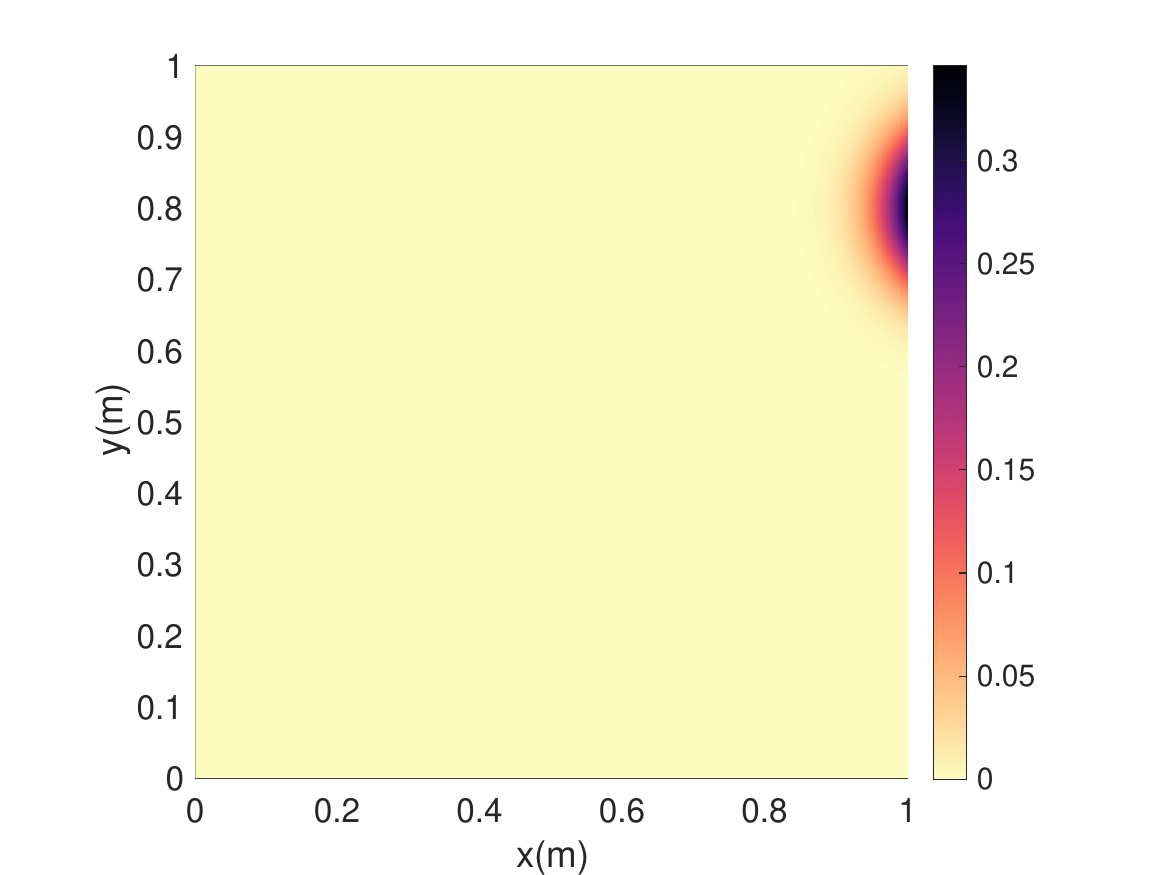}
        \caption{t=0.6s}
    \end{subfigure}

    \vspace{0.5em}  

    \caption{Velocity profile of the Gaussian vortex leaving the domain at different time.}
    \label{GaussianVortex}
\end{figure}

\begin{figure}[h!]
            \centering
            \includegraphics[width=0.7\textwidth]{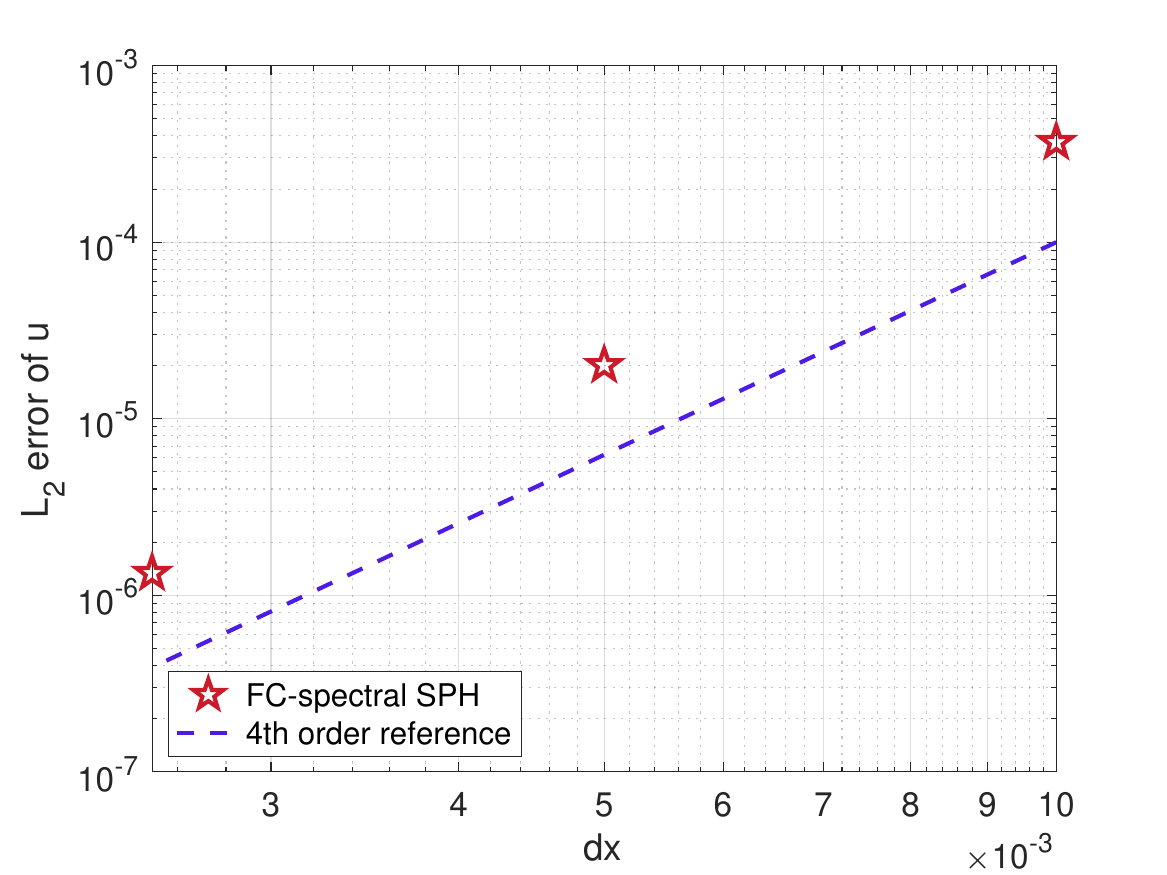}
            \caption{Convergence rate of $u$ in the Gaussian vortex convection case. }
            \label{L2GaussianVortex}
            \end{figure}

This non-periodic domain is extended by the aforementioned Fourier continuation, and then the spectral SPH scheme is implemented to approximate the velocity gradient. Figure \ref{GaussianVortex} presents the velocity profile of the Gaussian vortex leaving the domain at three different times. This illustrates the smooth propagation of the vortex as it leaves the computational domain. Convergence behaviour is shown in Figure \ref{L2GaussianVortex}, in which the $4th$ order accuracy is consistently achieved from the G4 SPH kernel and the high-order FC scheme. It is worth noting that imposing inflow-outflow boundary conditions within a purely spectral SPH framework is inherently challenging due to the global nature of the Fourier and trigonometric basis functions.

\subsection{Poiseuille flow}
\begin{figure}[h!]
            \centering
            \includegraphics[width=0.7\textwidth]{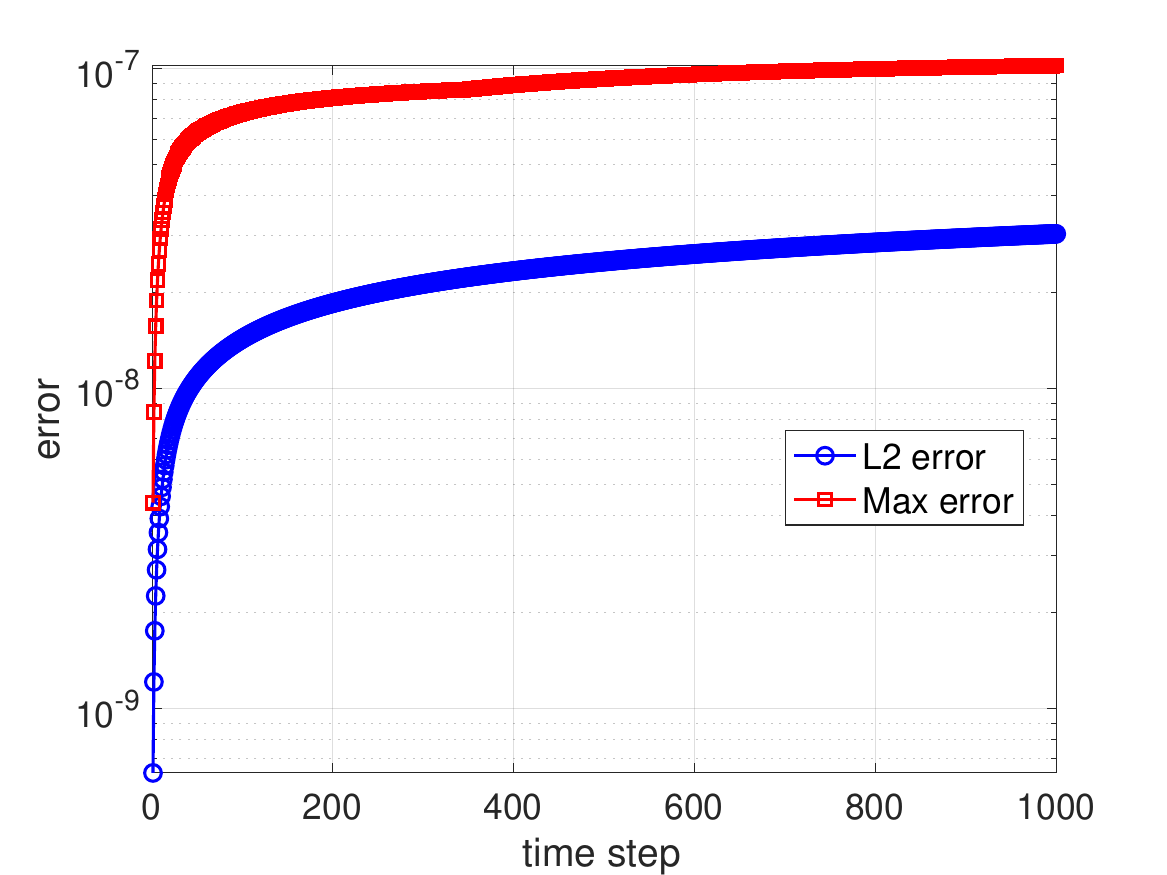}
            \caption{The $L_2$ evolution of the velocities in the Poiseuille flow case.}
            \label{PoiseuilleL2evolution}
            \end{figure}

\begin{figure}[h!]
            \centering
            \includegraphics[width=0.7\textwidth]{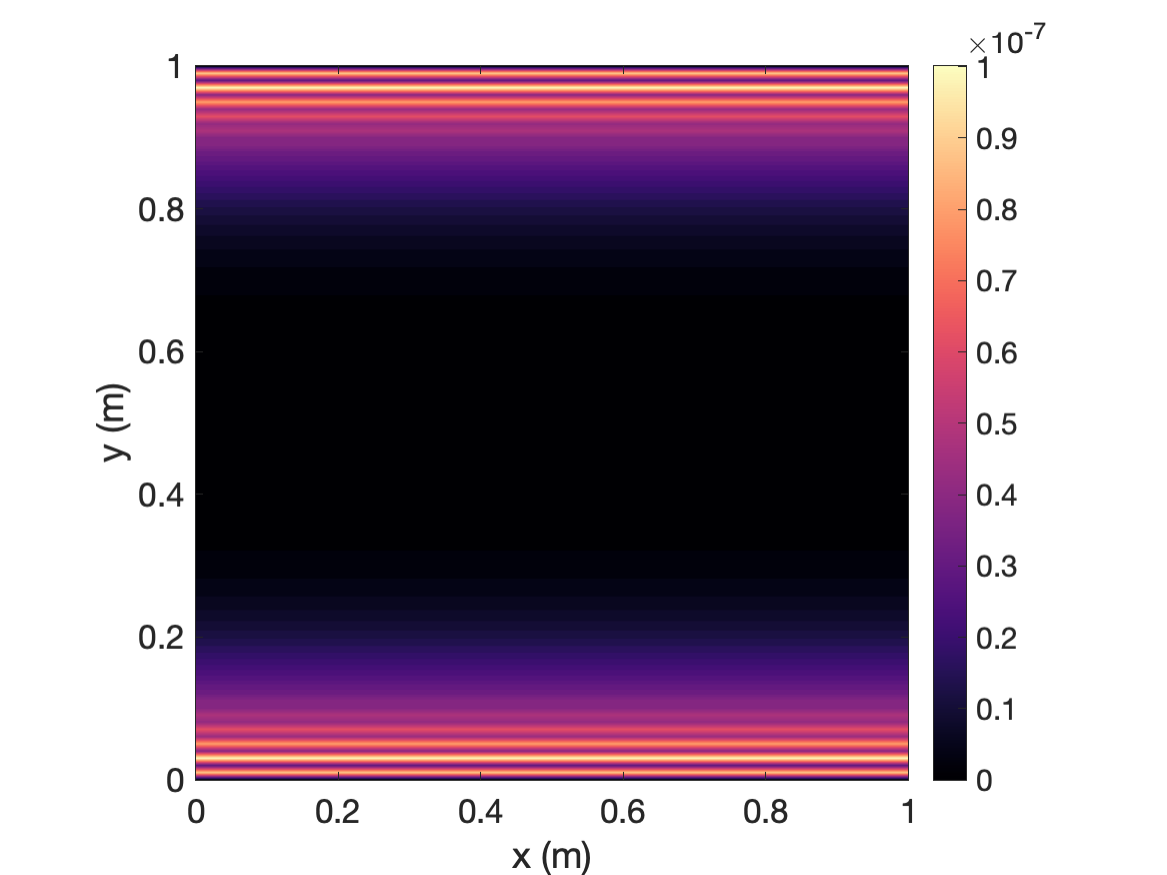}
            \caption{Pressure profile at $t=1.0s$ of the Poiseuille flow case.}
            \label{PoiseuilleP}
            \end{figure}

\begin{figure}[h!]
            \centering
            \includegraphics[width=0.7\textwidth]{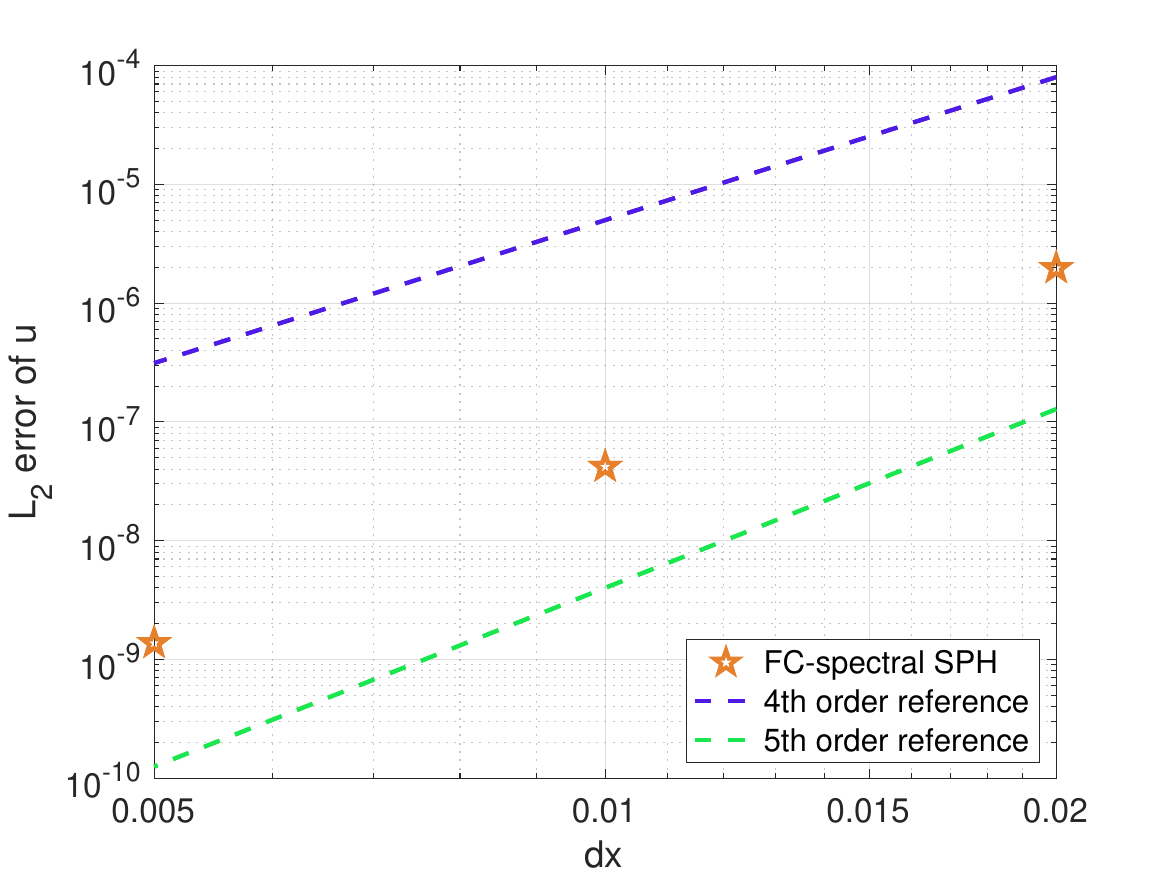}
            \caption{Convergence rate at $t=1$ s of the Poiseuille flow case.}
            \label{PoiseuilleConvergence}
            \end{figure}

From this section, the performance of the proposed FC-based spectral ISPH scheme is now examined within the framework of the full incompressible Navier-Stokes solver. As a simple benchmark validation case, Poiseuille flow is employed to assess the scheme. The computational configuration consists of a channel with periodic boundary conditions imposed in the streamwise direction and no-slip boundary conditions enforced in the wall-normal direction. The channel height is set to $H=1$ m, and the flow is driven by a spatially uniform body force $F_x$ applied in the streamwise direction. Under these conditions, the steady-state velocity profile is governed by the analytical solution shown in Equation (\ref{channelanalytical}):

 \begin{equation}\label{channelanalytical}
        u(y)=\frac{F_x}{2\nu}y(1.0-y).
     \end{equation}
All simulations are initialised with the analytical velocity profile at a Reynolds number of Re=100 and all the cases below are run for $t$ =$1$ s at which point the $L_2$ norm of the velocities field has effectively reached a plateau as shown in Figure \ref{PoiseuilleL2evolution}.
Figure~\ref{PoiseuilleP} presents the pressure field at $t = 1$ s for the Poiseuille flow case. The pressure distribution exhibits a spatially 
uniform profile in the streamwise direction, which is consistent with the periodic boundary conditions in this direction. The magnitude of the pressure field is of the order $\mathcal{O}(10^{-7})$. This confirms that the pressure fluctuations are negligibly small and the flow is driven by the imposed body force alone. Figure~\ref{PoiseuilleConvergence} presents the convergence rate of the FC-spectral SPH scheme. Notably, the scheme achieves a convergence rate that exceeds 4th-order accuracy. This super-convergence behaviour is attributed to the fact that the Poiseuille analytical 
solution is a quadratic polynomial, whose higher-order derivatives vanish 
identically. This might allow the scheme to achieve a convergence rate beyond the 4th-order from the G4 Gaussian kernel function. Meanwhile, the negligibly small pressure field ($\mathcal{O}(10^{-7})$) also eliminates pressure-velocity coupling errors, which also helps with the accuracy results.

\subsection{Plane Couette flow}
\begin{figure}[h!]
            \centering
            \includegraphics[width=0.7\textwidth]{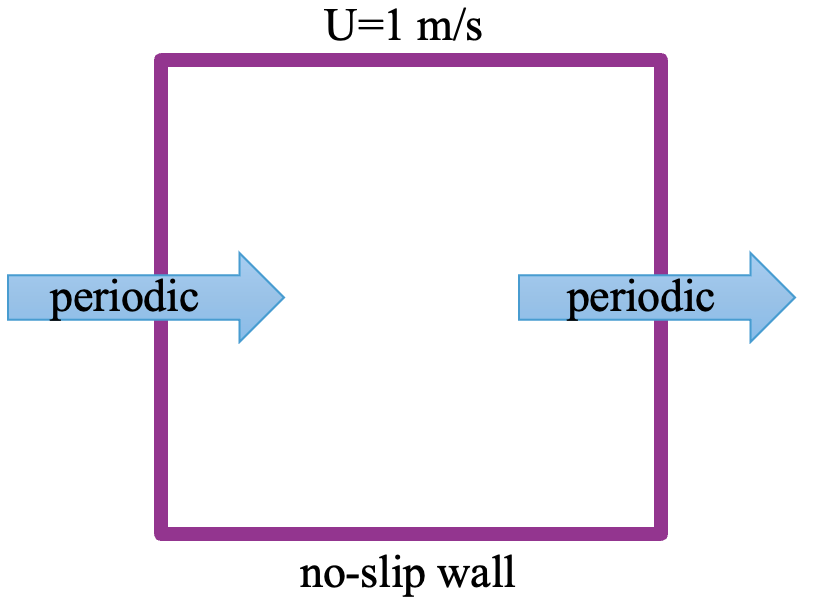}
            \caption{Boundary condition of the plane Couette flow.  }
            \label{CouetteGeo}
            \end{figure}

\begin{figure}[h!]
            \centering
            \includegraphics[width=0.7\textwidth]{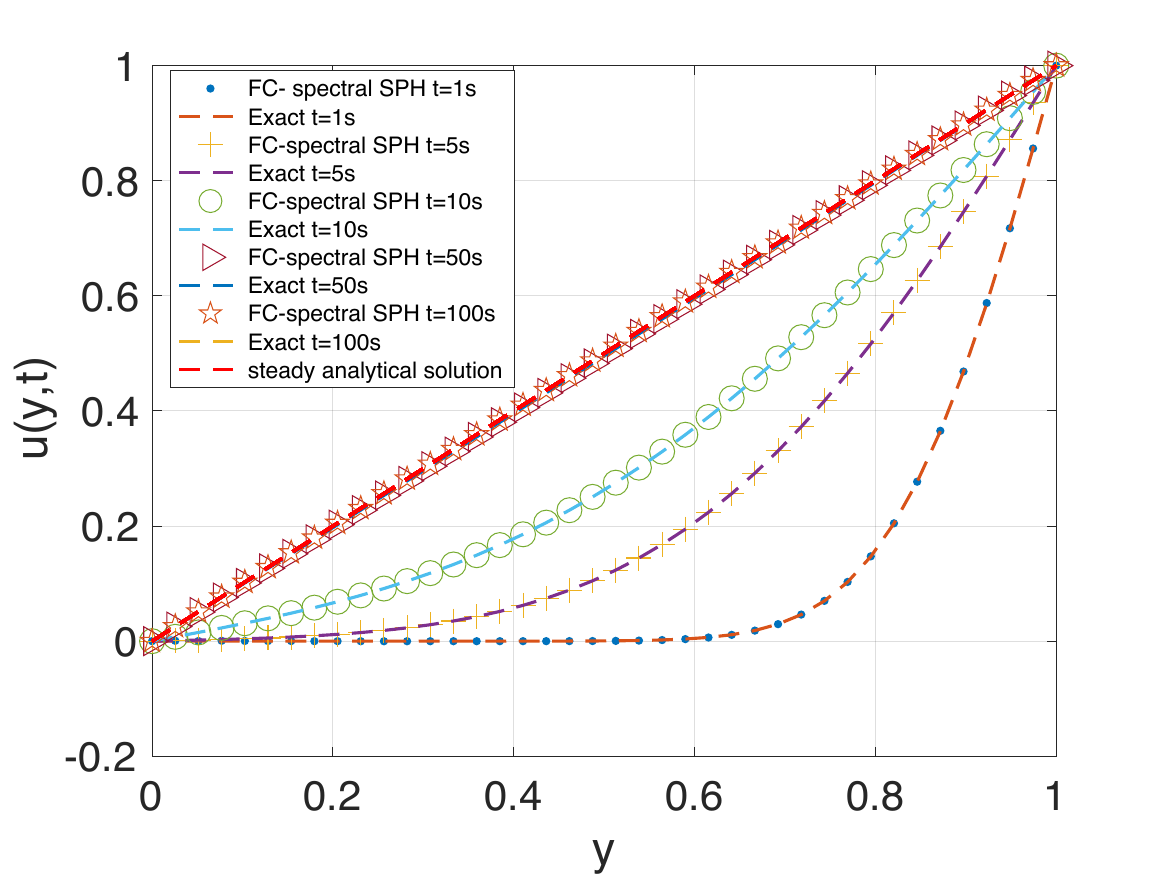}
            \caption{Comparison of the analytical transient velocity solution and the velocity obtained from the FC-based spectral ISPH solver. }
            \label{uCouette}
            \end{figure}

The previous test case validates the capability of the scheme to simulate the homogenous no-slip boundary condition.  In order to demonstrate that the FC scheme can be applied generally, the plane Couette flow is used as a test case, where the boundary conditions are $u=0$ at $y=0$ (stationary wall) and $u=1$ m/s at $y=1$ (moving wall), as shown in Figure \ref{CouetteGeo}. Such boundary condition setting represents a case where the velocity differs between the upper and lower walls. The simulation starts from zero velocity and is advanced in time until the velocity field $u$ converges to the analytical steady-state solution of plane Couette flow $u(y)=y$. Before reaching steady state, velocity profile $u$ is compared to the transient analytical solution at time $t$:
 \begin{equation}\label{CouetteTransient}
    u(y,t) = U_0 y + \frac{2U_0}{\pi} \sum_{n=1}^{\infty} \frac{(-1)^n}{n} \sin(n\pi y) \, e^{-\nu n^2 \pi^2 t},
\end{equation}
where $u(y,t) = U_0 y$ is the steady state analytical velocity profile. Such comparison is illustrated in Figure \ref{uCouette}, from which excellent agreement is observed between the FC-spectral SPH results and the analytical solution at all time instances, with the velocity profile evolving from the initial quiescent state and converging to the steady-state Couette profile.

It is noteworthy that, for a spectral SPH scheme without Fourier Continuation, the asymmetric boundary condition in 
the $y$ direction poses a difficulty, as the spectral transform is global. An appropriate splitting method \cite{Boyd2001,SplittingWang} must be adjusted to decompose the velocity into the homogeneous part and a simple function satisfying the boundary conditions. The FC-based spectral SPH scheme, however, requires no such treatment. Since the FC algorithm extends any non-periodic function to a smooth periodic one regardless of the prescribed boundary values, the asymmetric Dirichlet conditions of plane Couette flow are handled naturally and directly in physical space, without any modification to the scheme. This demonstrates the generality of the proposed method for arbitrary wall boundary conditions.

\subsection{Vortex dipole rebound from the wall}
\label{domainExtensionDipoleSection}
The vortex dipole rebound from a wall is chosen here as a validation benchmark for the FC-based spectral ISPH scheme. This test case has been widely used to evaluate the accuracy of various numerical approaches, such as pseudospectral and finite difference methods \cite{CLERCX2006245}, the volume penalisation immersed boundary method \cite{KEETELS2007919}, and the compact finite difference method \cite{LAIZET}. As noted in \cite{CLERCX2006245}, the vortex--wall interaction, despite its simple configuration, produces complex flow physics. When the dipole approaches the wall, a secondary vortex of opposite sign is generated within the boundary layer, this structure interacts with the primary dipole, leading to vortex rebound, additional small-scale vorticity, and enhanced energy dissipation. Consequently, resolving the full vortex structure requires very high spatial resolution. This test case is conducted to validate the proposed no-slip wall boundary condition, as well as the capability of the scheme to accurately resolve the complex vortex structures. Two counter-rotating vortices are initialized at $(x_1,y_1)=(-0.1,0)$ and $(x_2,y_2)=(0.1,0)$. The computational domain is defined as  $[-1,1]\times[-1,1]$, where both $x$ and $y$ directions are imposed with no-slip boundary conditions. The vorticity of each monopole is defined identically as in \cite{CLERCX2006245}:
\begin{equation}\label{vorticity}
\omega_0=\omega_e\left(1-\left(\frac{r}{r_0}\right)^2\right)e^{\left(-\left(\frac{r}{r_0}\right)^2\right)}.
\end{equation}
Here $r=0.1$ is the monopole radius and $\omega_e$ denotes the extremum vorticity. Its magnitude is set to $299.528$ to so that the initial value $ E(0)$ of the total kinetic energy $E(t)=\frac{1}{2}\int_{-1}^{1}\int_{-1}^{1}\textbf{u}^2(\textbf{x},t)dxdy$ equals $2$. The initial velocity $\textbf{u}=(u_0,v_0)$ is therefore, determined from the prescribed vorticity: 

\begin{equation}\label{u0Vortex}
u_0 =
-\frac{1}{2}\lvert \omega_e\rvert (y - y_1)
  e^{\left[-\left(\frac{r_1}{r_0}\right)^{2}\right]}
+\frac{1}{2}\lvert \omega_e\rvert (y - y_2)
 e^{\left[-\left(\frac{r_2}{r_0}\right)^{2}\right]},
\end{equation}

\begin{equation}\label{v0Vortex}
v_0 =
 \frac{1}{2}\lvert \omega_e\rvert (x - x_1)
  e^{\left[-\left(\frac{r_1}{r_0}\right)^{2}\right]}
-\frac{1}{2}\lvert \omega_e\rvert (x - x_2)
  e^{\left[-\left(\frac{r_2}{r_0}\right)^{2}\right]},
\end{equation}
where $r_1 = \sqrt{(x - x_1)^2 + (y - y_1)^2}$
and $r_2 = \sqrt{(x - x_2)^2 + (y - y_2)^2}$.
The dipole rebound is simulated at two Reynolds numbers: 100 and 625.

\subsubsection{Re=100}
\begin{figure}[h!]
    \centering
    \begin{subfigure}{.36\textwidth}
        \centering
        \includegraphics[width=\linewidth]{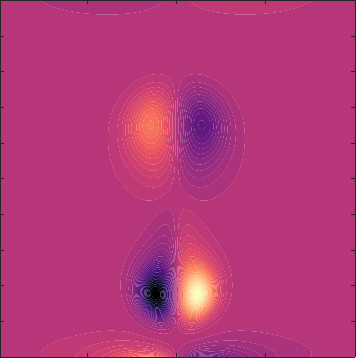}
        \caption{FC-spectral ISPH, $t=5$ s}
    \end{subfigure}
    \begin{subfigure}{.39\textwidth}
        \centering
        \includegraphics[width=\linewidth]{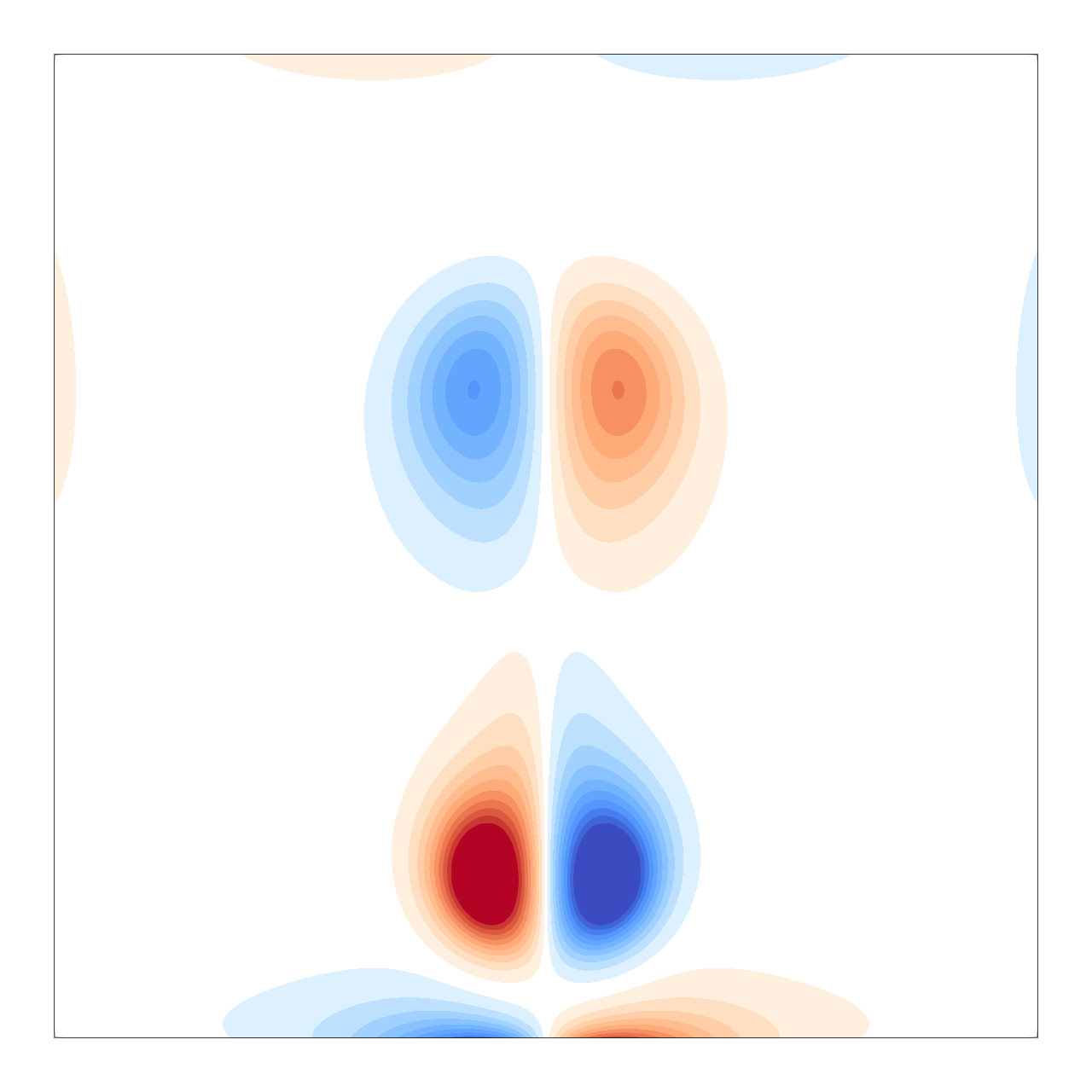}
        \caption{Spectral element, $t=5$ s}
    \end{subfigure}

    \vspace{0.5em}

    \begin{subfigure}{.36\textwidth}
        \centering
        \includegraphics[width=\linewidth]{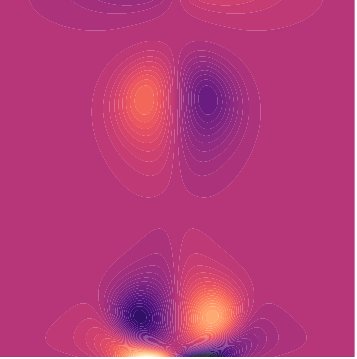}
        \caption{FC-spectral ISPH, $t=10$ s}
    \end{subfigure}
    \begin{subfigure}{.39\textwidth}
        \centering
        \includegraphics[width=\linewidth]{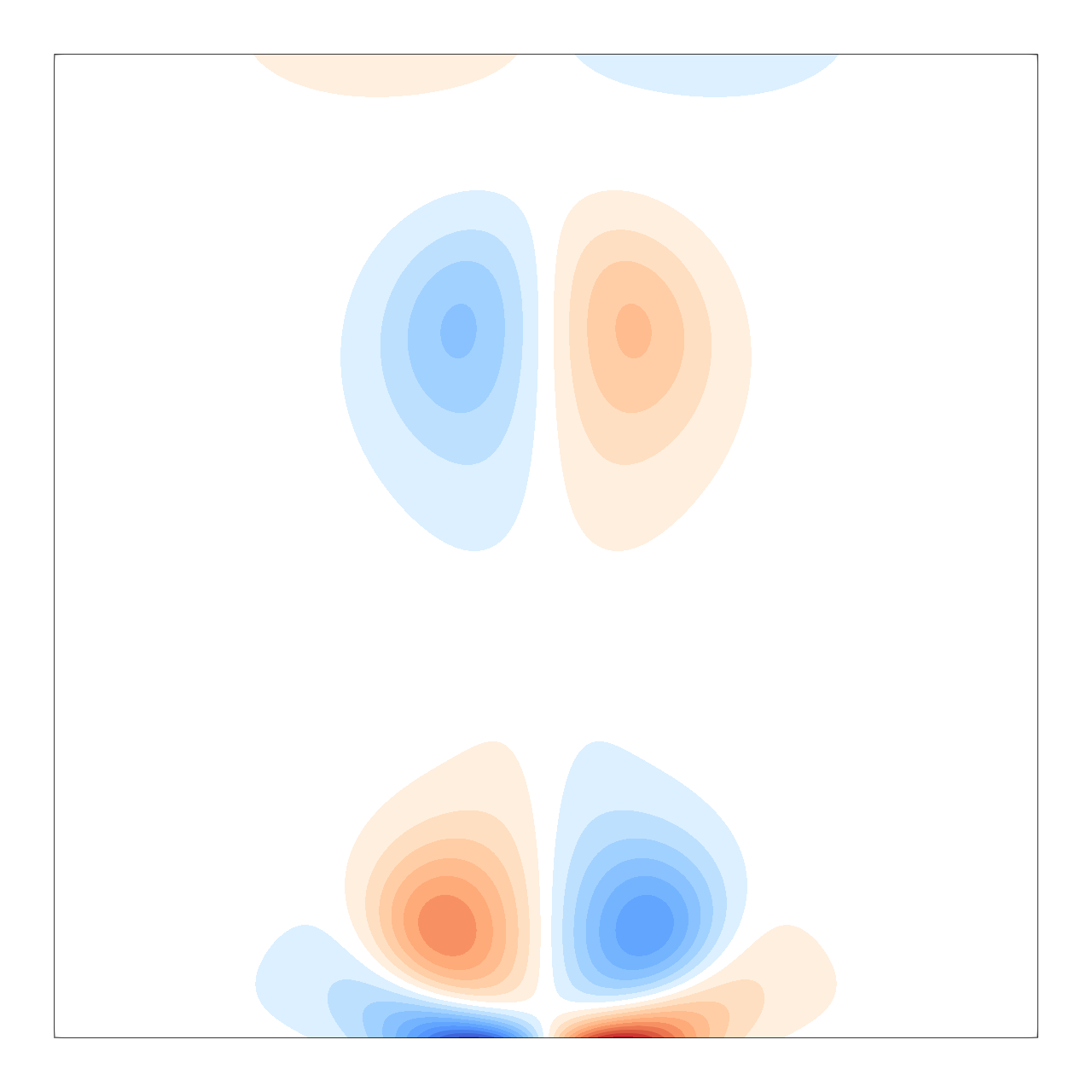}
        \caption{Spectral element, $t=10$ s}
    \end{subfigure}

    \caption{Vorticity contours of the vortex dipole at $Re=100$, comparing the FC-spectral ISPH scheme (left) with the spectral element reference solution (right) at $t=5$ s (a,b) and $t=10$ s (c,d).}
    \label{fig:dipole_vorticity}
\end{figure}

For Re=100, we choose a marginal resolution of $161 \times 161$ particles, with $dt=1.25\times10^{-3}$. The large viscosity at this Reynolds number suppresses fine-scale structures, thus this resolution is sufficient to capture the rebound dynamics. To benchmark the proposed method against a well-established high-order numerical framework, the same test case with identical parameters is implemented in the open-source $\textit{hp}$/spectral element (SEM) solver Nektar++ \cite{Nektar++}. This comparison provides a rigorous reference for assessing the accuracy and consistency of the proposed approach. In this configuration, the incompressible unsteady Navier–-Stokes equations are discretised using a continuous Galerkin formulation with a maximum polynomial expansion order of $P=7$ to provide high-order spatial accuracy. All the results from Nektar++ are obtained with a resolution of 78400 nodes. Figure \ref{fig:dipole_vorticity} shows the vorticity results of $t=5$ s and $t=10$ s. At $t=10$ s, following the rebound from the wall, the secondary vorticity structures generated near the boundary are also well resolved by the FC-based ISPH scheme.

\begin{figure}[h!]
    \centering
   
    \begin{subfigure}{.48\textwidth}
        \centering
        \includegraphics[width=\linewidth]{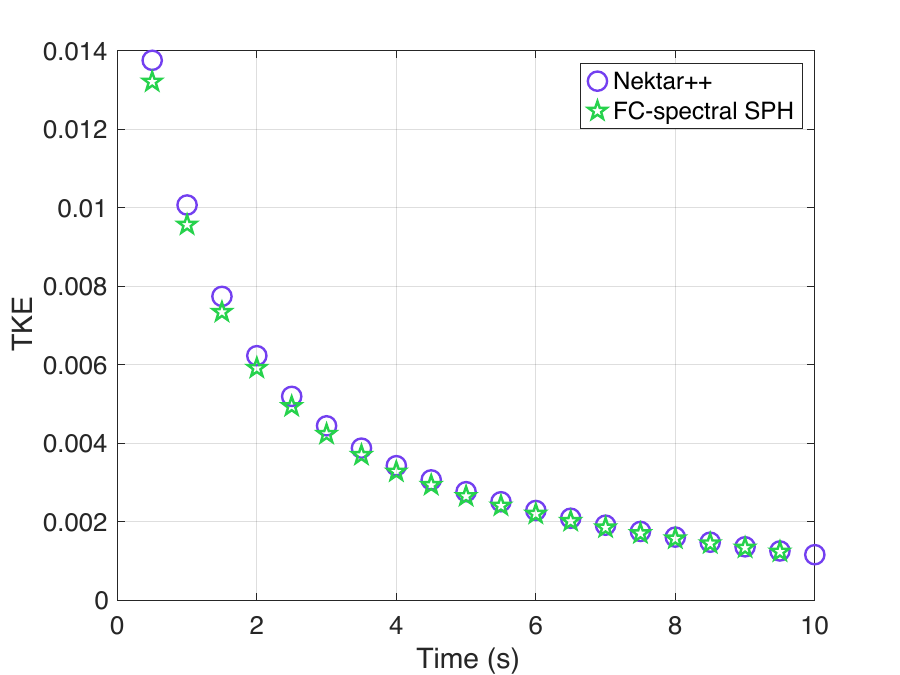}
        \caption{TKE}
    \end{subfigure}
    \begin{subfigure}{.48\textwidth}
        \centering
        \includegraphics[width=\linewidth]{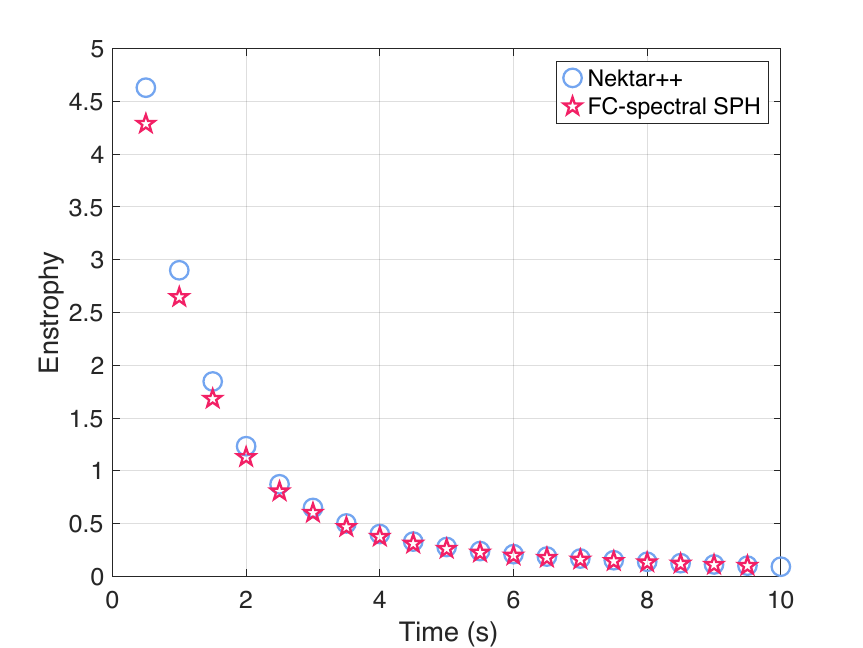}
        \caption{Enstrophy}
    \end{subfigure}

    \vspace{0.5em}  

    \caption{Comparison of the TKE and the enstrophy for the FC-based spectral ISPH scheme and the spectral element method results from Nektar++ at t=$10$ s for the dipole rebound at $Re=100$.}
    \label{Re_100_TKE_enstrophy}
\end{figure}

    Quantitatively, Figure \ref{Re_100_TKE_enstrophy} shows the temporal evolution of the total kinetic energy (TKE) E(t) and enstrophy Z(t) of both schemes defined in the following equations,
     \begin{equation} \label{E(t)}
         E(t)=\frac{1}{2}\int_\Omega (u^2+v^2)d\Omega,
     \end{equation} 
      \begin{equation} \label{Z(t)}
         Z(t)=\frac{1}{2}\int_\Omega \omega^2d\Omega,
     \end{equation} 
     where $\Omega$ is the computational domain and $\omega$ is the vorticity. Both schemes accurately capture the global decay of total kinetic energy and the correct monotonic decay of enstrophy.  A close agreement is observed between the FC-based spectral ISPH scheme and the SEM. Before $t=3$ s, the FC-based ISPH scheme prediction is slightly lower than the SEM, this is due to the inherent smoothing from the SPH kernel functions.

     \subsubsection{Re=625}
\begin{figure}[h!]
    \centering
    \begin{subfigure}{.30\textwidth}
        \centering
        \includegraphics[width=\linewidth]{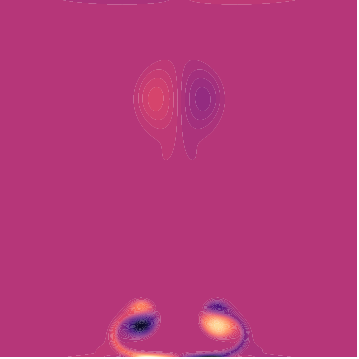}
        \caption{spectral ISPH ($200\times200$)}
    \end{subfigure}
    \hspace{0.5em}
    \begin{subfigure}{.30\textwidth}
        \centering
        \includegraphics[width=\linewidth]{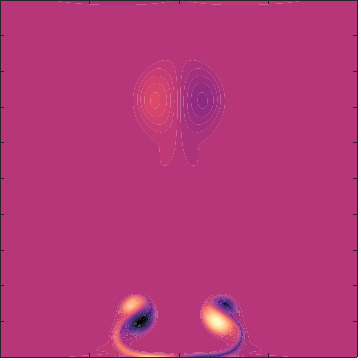}
        \caption{spectral ISPH ($600\times600$)}
    \end{subfigure}
    \hspace{0.5em}
    \begin{subfigure}{.30\textwidth}
        \centering
        \includegraphics[width=\linewidth]{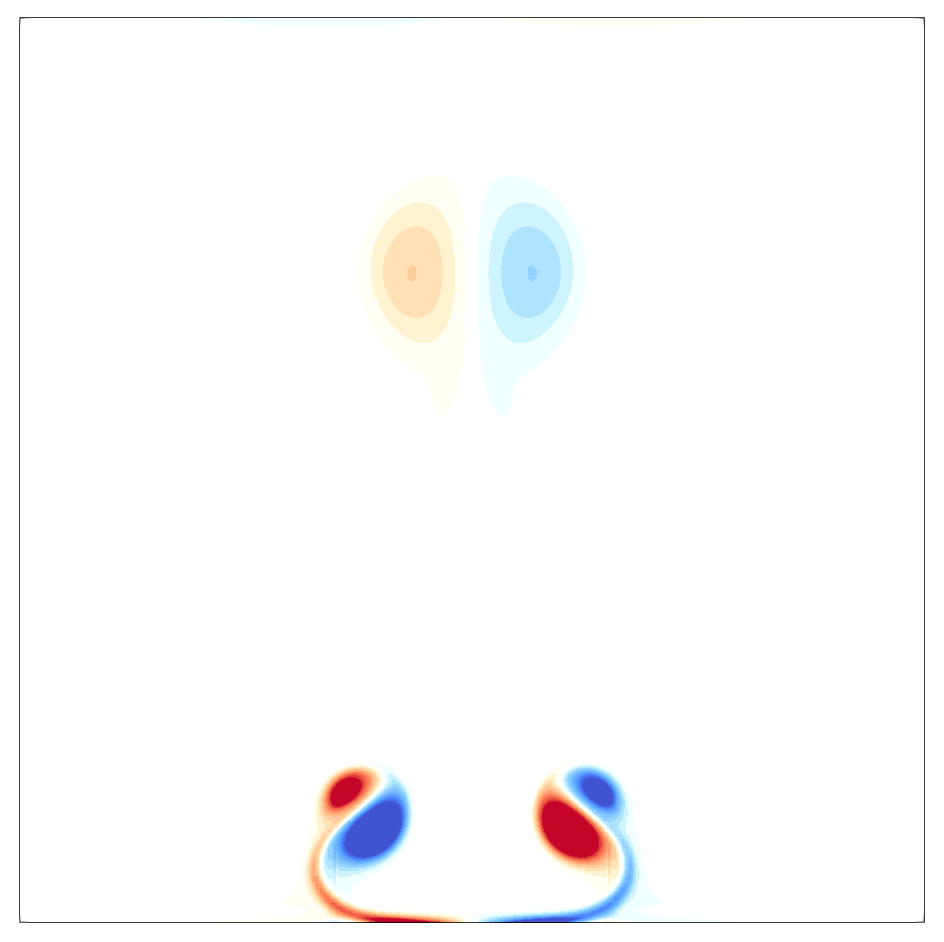}
        \caption{Spectral element}
    \end{subfigure}
    \vspace{0.5em}
    \begin{subfigure}{.30\textwidth}
        \centering
        \includegraphics[width=\linewidth]{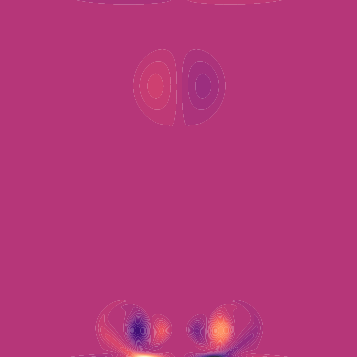}
        \caption{spectral ISPH ($200\times200$)}
    \end{subfigure}
    \hspace{0.5em}
    \begin{subfigure}{.30\textwidth}
        \centering
        \includegraphics[width=\linewidth]{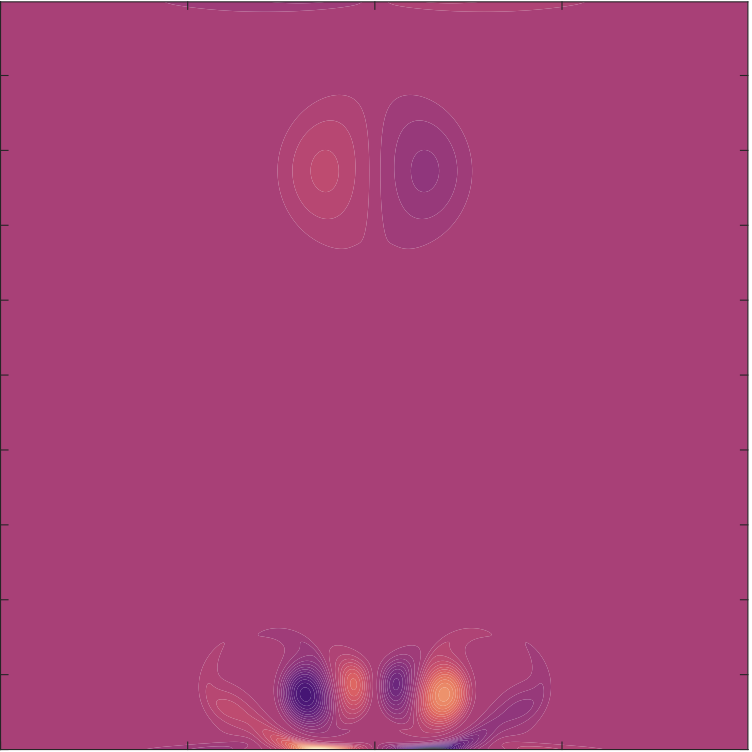}
        \caption{spectral ISPH ($600\times600$)}
    \end{subfigure}
    \hspace{0.5em}
    \begin{subfigure}{.30\textwidth}
        \centering
        \includegraphics[width=\linewidth]{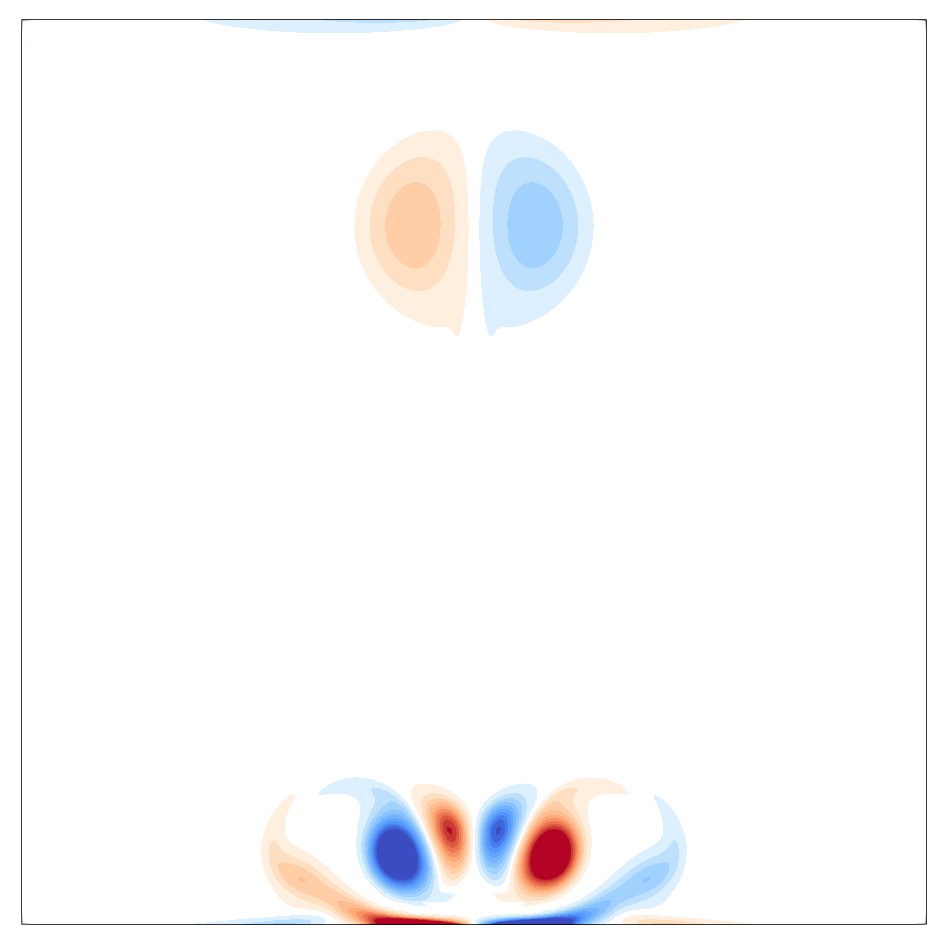}
        \caption{Spectral element}
    \end{subfigure}
    \vspace{0.5em}
    \begin{subfigure}{.30\textwidth}
        \centering
        \includegraphics[width=\linewidth]{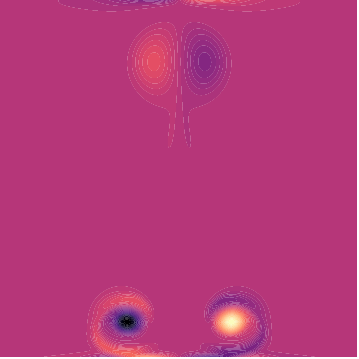}
        \caption{spectral ISPH ($200\times200$)}
    \end{subfigure}
    \hspace{0.5em}
    \begin{subfigure}{.30\textwidth}
        \centering
        \includegraphics[width=\linewidth]{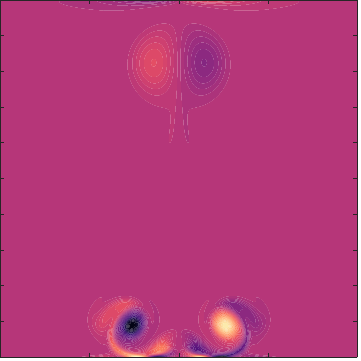}
        \caption{spectral ISPH ($600\times600$) }
    \end{subfigure}
    \hspace{0.5em}
    \begin{subfigure}{.30\textwidth}
        \centering
        \includegraphics[width=\linewidth]{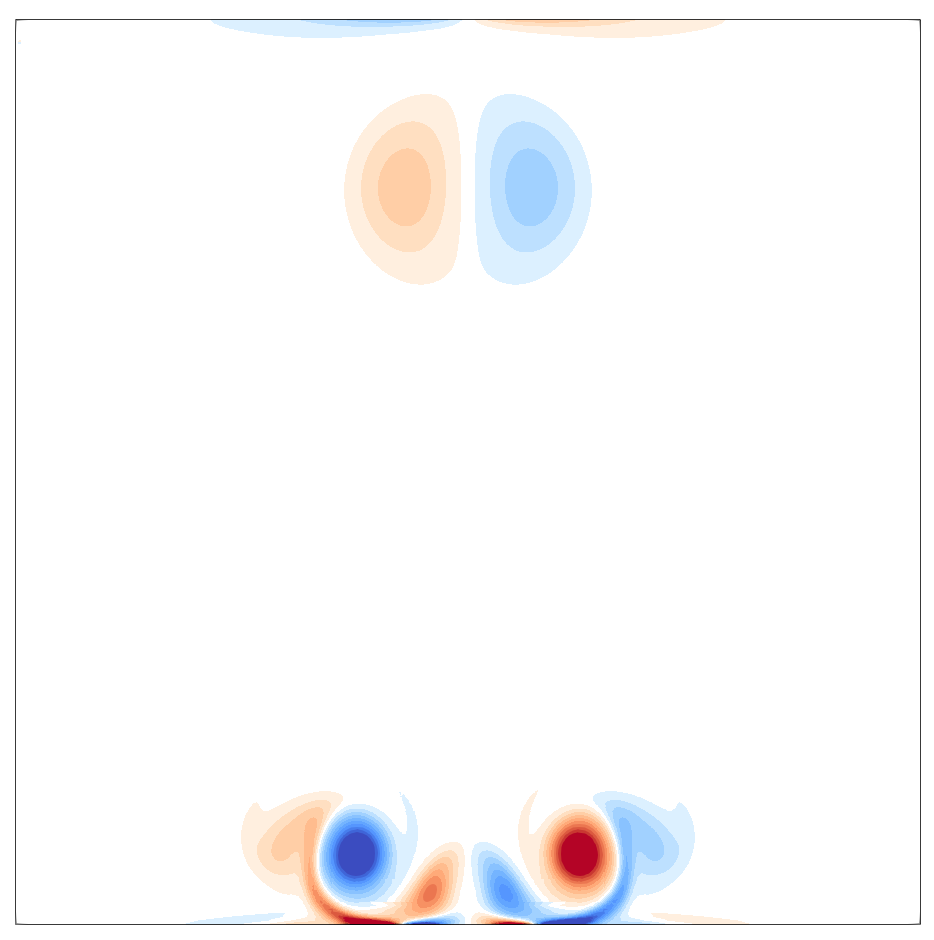}
        \caption{Spectral element}
    \end{subfigure}
    \caption{Vorticity contours of the vortex dipole at $Re=625$, comparing the FC-spectral ISPH scheme with $200\times 200$ particles (left), FC-spectral ISPH scheme with $600\times 600$ particles (middle), with the spectral element solution (right) at $t=0.5$ s (a,b,c), $t=0.65$ s (d,e,f), and $t=0.8$ s (g,h,i).}
    \label{Re625Vorticity}
\end{figure}

\begin{figure}[h!]
            \centering
            \includegraphics[width=0.7\textwidth]{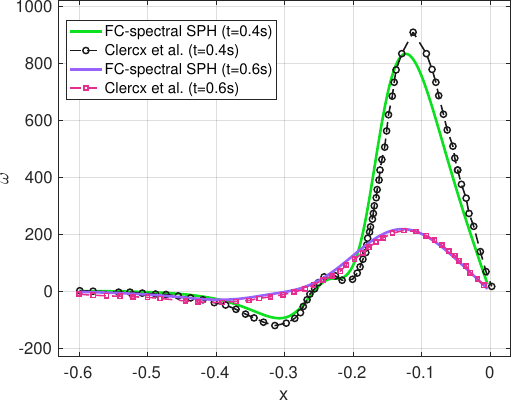}
            \caption{Vorticity at $y=-1$ at different $t$ from the FC-spectral ISPH scheme and the Chebyshev pseudospectral solver in \cite{CLERCX2006245}. }
            \label{Re_625_vorticityClercx}
            \end{figure}

     To evaluate the performance of the FC-spectral ISPH solver for high Reynolds number and thin shear layers, we run the dipole rebound case under Re=625.  For qualitative comparisons, Figure \ref{Re625Vorticity} shows the vorticity isocontour obtained by the proposed FC-spectral ISPH scheme and the spectral element methods at different time instances. Both the FC-spectral ISPH scheme with a resolution of $600 \times 600$ and the SEM show a sensible prediction of the vortex dynamics after the collision: secondary vortices detach from the wall and interact with the primary dipole, producing the characteristic rebound and subsequent roll-up observed in the literature. The contours at all three time instances are in close agreement between the two methods. This confirms that the FC-spectral scheme accurately resolves the thin boundary layers and vorticity generation at the no-slip wall. At the coarser resolution of $200\times 200$, the overall vortex shape is preserved, but the secondary structures appear more diffused. This  indicates that this resolution is insufficient to fully resolve the fine-scale dynamics at $Re=625$. Quantitatively, Figure \ref{Re_625_vorticityClercx} shows the vorticity distribution along the no-slip wall $y=-1$ at $t=0.4$ s and $t=0.6$ s. We compare the values with that reported in \cite{CLERCX2006245} by a Cherbyshev pseudospectral methods. At $t=0.4$ s, the FC-spectral ISPH scheme captures the sharp vorticity peak near $x=-0.2$, though the peak magnitude is slightly underpredicted compared to the reference. The overall shape of the wall vorticity distribution is well reproduced. At $t=0.6$ s, both the peak location and magnitude closely match the reference solution. The slight discrepancy at $t=0.4$ s can be attributed to the large vorticity gradients generated during the initial wall collision, which are the most demanding to resolve. As reported in Clercx et al \cite{CLERCX2006245}, at least a resolution of $768$ mesh points in each direction is needed with finite difference methods to fully resolved the vortex shear layer. As the flow evolves, the FC-spectral scheme recovers good agreement with the reference. This case validates the capability of the FC-spectral ISPH scheme in resolving complex vortex--wall interactions at different Reynolds numbers, where thin shear layers and secondary vortex detachment pose significant challenges for the numerical schemes to accuracely capture the fluid physics.

\clearpage
\section{Conclusion} \label{Conclusion}




This paper proposes a novel Fourier continuation (FC)-based spectral 
incompressible SPH scheme. The scheme extends the existing spectral ISPH framework 
to support high-order wall boundary conditions by applying FC to overcome the periodicity requirement. The continuation achieves $C^p$ smoothness via high-order polynomial fitting, extrapolation and blending, to recover the high-order convergence of the spectral SPH operators. With an extension length of $25\%$ 
of the fluid domain, the additional computational cost remains 
acceptable. Compared with direct even/odd domain extension of the whole fluid domain, the proposed 
scheme achieves superior accuracy and efficiency.


The scheme is validated against classical CFD benchmarks with no-slip, Dirichlet, and Neumann boundary conditions, inflow--outflow vortex convection, and vortex--wall interaction at different Reynolds numbers. Quantitative and qualitative results confirm high-order convergence and the accurate capture of vortex dynamics. 
Whilst showing satisfying results in these benchmarks, the current scheme has some limitations. Firstly, the Fourier continuation introduced in this work is restricted to regular and smooth computational domains. Moreover, the current scheme rely on the Cartesian particle distribution. This limits it application on complex geometries. These two limitations will be addressed in future work by adopting the two-dimensional FC framework of Bruno et al.~\cite{2DFC} for general smooth domains, which will ultimately enable a fully Lagrangian spectral ISPH solver for complex real-world applications.

\section*{Acknowledgment}
The authors would like to thank the computational facility at the University of Manchester. Author Meixuan Lin would like to gratefully acknowledge the Department of Mechanical, Aerospace and Civil Engineering at the University of Manchester for providing funding for the Ph.D research project in which this work is developed.

\appendix
\section{Derivation of the 8th order Gaussian kernel $W_{\text{G8}}$}
\label{appendixA}

The 8th-order Gaussian SPH kernel function G8 is derived here. G8 is implemented in the vortex dipole rebound tese case at Re=625. The derivation follows the construction of 4th and 6th order Gaussian kernel in the work of Lind et al. \cite{LIND2016JCP}. The SPH integral approximation of a function $A$ at point $\textbf{r}_i$ is given by the convolution of $A$ with the kernel function $W(\textbf{r}_i-\textbf{r})$:

\begin{equation}
A(\textbf{r}_i)\approx\int_\Omega A(\textbf{r})W(\textbf{r}_i-\textbf{r})dV.
    \label{StevenA(r)}
\end{equation}
Applying Taylor series expansion of $A(\textbf{r}_i)$ and substituting into Equation (\ref{StevenA(r)}) leads to the following formulation of $A(\textbf{r}_i)$:

\begin{equation}
A_i=A_i\int WdV+h\int A_i'qWdV+\frac{h^2}{2}\int A_i''q^2WdV+\frac{h^3}{6}\int A_i'''q^3WdV+\mathcal{O} (h^4),
    \label{StevenTaylor}
\end{equation}
where $q=\frac{r}{h}$ is a non-dimensional number in the kernel function, $A_i$ is short for $A(\textbf{r}_i)$, and $V$ is the volume of the particle. As pointed out in \cite{LIND2016JCP}, the formulation of the high-order kernels can be generalised by removing all the even moments:
\begin{equation}
W_{2n}=\sum_{m=0}^{n-1} (A_mq^{2m})W_{\text{G2}}.
    \label{generalKernel}
\end{equation}\

Using $I_{n}$ to denote the $n$th moment $\int q^{2n} \omega \, dV$ of the 2nd order
Gaussian, the conditions become:

\begin{equation}
\sum_{m=0}^{3} A_m I_{m+j} = \delta_{j0}, \quad j = 0, 1, 2, 3
\end{equation}

which yields the $4 \times 4$ linear system:

\begin{equation}
\begin{pmatrix} I_0 & I_1 & I_2 & I_3 \\ I_1 & I_2 & I_3 & I_4 \\ 
I_2 & I_3 & I_4 & I_5 \\ I_3 & I_4 & I_5 & I_6 \end{pmatrix} 
\begin{pmatrix} A_0 \\ A_1 \\ A_2 \\ A_3 \end{pmatrix} = 
\begin{pmatrix} 1 \\ 0 \\ 0 \\ 0 \end{pmatrix}
\end{equation}
In 2D, the moments of the Gaussian base kernel evaluate to $I_k = k!$, and solving this gives $A_0 = 4$, $A_1 = -6$, $A_2 = 2$, $A_3 = -\tfrac{1}{6}$, 
which is  the G8 kernel:

\begin{equation}
W^{\text{G8}}(q) = \alpha_{\text{G8}} \left(4 - 6q^2 + 2q^4 - 
\frac{q^6}{6}\right) e^{-q^2},
\end{equation}
where $\alpha_{\text{G8}} = 1/(\pi h^2)$ in 2D.


\bibliographystyle{elsarticle-num}
\bibliography{references}






\end{document}